\documentclass[11pt,a4paper]{article}
\usepackage{jcappub}
\usepackage{bm}
\usepackage{bbm}
\usepackage{amsmath}
\usepackage{etoolbox}
\usepackage{float}

\renewcommand\[{\left[}

\newcommand{\bk}{{\bf k}}

\newcommand{\sE}{{\sf E}}
\newcommand{\feq}{f_{\rm eq}}

\long\def\exclude#1{}

\usepackage{etoolbox}

\makeatletter
\gdef\@fpheader{}
\makeatother

\begin{document}

\makeatletter
\makeatother

\title{Radiative transfer in stars by feebly interacting bosons}

\author[a,b]{Andrea Caputo,}
\author[c]{Georg Raffelt,}
\author[d,e]{and Edoardo Vitagliano}

\affiliation[a]{School of Physics and Astronomy, Tel-Aviv University, \\ Tel-Aviv 69978, Israel}
\affiliation[b]{Department of Particle Physics and Astrophysics, Weizmann Institute of Science, \\Rehovot 7610001, Israel}
\affiliation[c]{Max-Planck-Institut f\"ur Physik (Werner-Heisenberg-Institut), \\F\"ohringer Ring 6, 80805 München, Germany}
\affiliation[d]{Department of Physics and Astronomy, University of California, Los Angeles, \\ 475 Portola Plaza, Los Angeles, CA 90095-1547, USA}
\affiliation[e]{Niels Bohr International Academy and DARK, Niels Bohr Institute, University of Copenhagen, Blegdamsvej 17, 2100, Copenhagen, Denmark}


\emailAdd{andreacaputo@mail.tau.ac.il}
\emailAdd{raffelt@mpp.mpg.de}
\emailAdd{edoardo@physics.ucla.edu}

\exclude{
\abstract{Starting from first principles, we calculate the luminosity of a supernova core in the form of new feebly-interaction bosons (FIBs) such as axions, axion-like particles (ALPs), dark photons, and others. As a function of interaction strength, we study the transition from the free-streaming to the trapping regime and the conceptual question of how volume emission turns to quasi-surface quasi-thermal emission from a ``FIB sphere.'' Our nominal result is a spherical volume-integral expression using the reduced absorption rate $\Gamma$ (that depends on radius and may depend on FIB energy) as the only particle-physics ingredient. Our result supersedes expressions and approximations found in the recent literature. As a specific example we consider ALP production based on the Primakoff process.}
}

\abstract{Starting from first principles, we study radiative transfer by new feebly-interacting bosons (FIBs) such as axions, axion-like particles (ALPs), dark photons, and others. Our key simplification is to include only boson emission or absorption (including decay), but not scattering between different modes of the radiation field. Based on a given distribution of temperature and FIB absorption rate in a star, we derive explicit volume-integral expressions for the boson luminosity, reaching from the free-streaming to the strong-trapping limit. The latter is seen explicitly to correspond to quasi-thermal emission from a ``FIB sphere'' according to the Stefan-Boltzmann law. Our results supersede expressions and approximations found in the recent literature on FIB emission from a supernova core and, for radiatively unstable FIBs, provide explicit expressions for the nonlocal (``ballistic'') transfer of energy recently discussed in horizontal-branch stars.}

\maketitle

\clearpage

\section{Introduction}

Dark sectors arising from physics beyond the standard model could provide explanations for various shortcomings of the standard model itself, including dark matter, neutrino masses, the baryon asymmetry, and the strong CP problem. One typical phenomenological consequence is the appearance of new, feebly-interacting bosons (FIBs) that can be experimentally searched and astrophysically or cosmologically constrained. One class of traditional arguments uses observational consequences of FIB emission from stars, an idea independently advanced by several groups in 1978 \cite{Mikaelian:1978jg,Dicus:1978fp,Vysotsky:1978dc,Sato:1978vy} when the Weinberg-Wilczek axion had been recognized as a consequence of the Peccei-Quinn solution of the strong CP problem. Ever since, the impact of many types of bosons in various astrophysical systems has been studied \cite{Raffelt:1996wa}, sometimes posing interesting conceptual questions about FIB production or propagation in stars. 

We here follow up one such case that has emerged in several recent studies of FIB production in supernova (SN) cores \cite{Chang:2016ntp, Chang:2018rso, Lucente:2020whw,Bollig:2020xdr,Caputo:2021rux,Caputo:2022mah, Croon:2020lrf}.
Actually the feeble interaction was taken strong enough to prevent free escape after production. In analogy to the SN ``neutrino sphere,'' the FIBs emerge from a decoupling region that is traditionally pictured approximately as a black surface for thermal FIB radiation according to the Stefan-Boltzmann (SB) law \cite{Burrows:1990pk}. The relevant temperature $T_{\rm SB}$ is taken to be that of the SN medium at a radius $R_{\rm SB}$ where the FIB optical depth is 2/3, and the radiating surface is $4\pi R_{\rm SB}^2$. We will see that this prescription is rather accurate, as physically it should be, but has evoked some doubts because clearly there is no hard surface of emission---the radiation must come from a shell with a geometric thickness corresponding to optical depth of around one.

Motivated by this question and doubts in the recent literature we take a fresh look at radiative transfer by FIBs that may or may not have a significant mass. In the diffusion limit, this problem was formulated a long time ago \cite{Raffelt:1988rx}, following the standard theory of radiative transfer by photons.\footnote{A free electronically available textbook is Rutten (2003) \cite{Rutten:2003}. It provides a fantastic annotated biblio\-graphy and references both to the early papers by Schuster, Schwarzschild, Eddington, Rosseland and Milne as well as to many textbooks, explaining their focus and relevance. For our work, we have mostly consulted the classic textbook by Mihalas (1978) \cite{Mihalas:1978} and Appendix~I of Shapiro and Teukolsky (1983) \cite{Shapiro:1983du}. See also Chapter~3, Sec.~3.4, of Refs.~\cite{ThorneBook,ThorneWeb} for some useful definitions of angular moments related to our Sec.~\ref{SubSec:Angular}.} Our focus here is to study explicitly the transition between the free-streaming and trapping (diffusion) limits, both in plane-parallel and spherical geometry. The latter is particularly interesting in a situation when the FIBs are unstable and deposit energy in regions far away from the compact emission volume, i.e., in a situation where the geometric extension of the ``stellar atmosphere'' is not much smaller, or even much larger, than the core radius of a SN or a horizontal-branch or red-giant star \cite{Lucente:2022wai}.

The main simplifying assumption, motivated by the boson interaction being ``feeble'', is to include only FIB absorption and emission from a medium in local thermal equilibrium, but not scattering between different FIB momenta or annihilation. In this case the only particle-physics ingredient is the ``reduced absorption rate'' $\Gamma_\omega$ as a function of FIB energy $\omega$, where $\Gamma_\omega$ is equivalent to the imaginary part of the FIB self-energy, that also depends on the local conditions of the medium such as temperature, density, and chemical composition. In the absence of scattering, the stationary FIB occupation number on a given ray, corresponding to a given mode $\bk$ of the FIB radiation field, can be expressed as an integral along this ray. Global solutions for plane-parallel or spherical geometries then follow as suitable superpositions of such single-ray solutions. In other words, for a given stationary stellar background model, the FIB radiation field is found from a quadrature. Explicit volume-integral expressions, notably in spherical geometry, are the main technical results of our paper. Based on $\Gamma_\omega(r)$ and $T(r)$ as functions of stellar radius, we thus provide integral expressions for the FIB luminosity $L_\omega(r)$. Taken at spatial infinity, $\int d\omega\,L_\omega(\infty)$ provides the total FIB luminosity, e.g., of a SN core. Moreover, one can find the energy loss or deposition at a given radius through the radial variation $dL_\omega(r)/dr$.

Solutions derived from a prescribed and stationary background model are only useful, of course, in a physical situation when the thermal timescale exceeds the dynamical one. If this is not the case, and if the diffusion limit does not apply, the full Boltzmann collision equation needs to be solved, a task that is of course the main numerical effort in core-collapse SN simulations concerning neutrino transport.

Radiative transport by neutrinos, despite their weak interaction, is a much more complicated task than our FIB treatment.
Neutrinos and antineutrinos of the electron and muon flavor can be absorbed and emitted by the medium through charged-current interactions, but neutral-current scatterings as well as annihilation and pair emission and absorption through bremsstrahlung and other processes occur on the same order of the coupling constant $G_{\rm F}^2$. Moreover, besides energy also lepton number of different flavors is being transported.

In principle, our exercises are straightforward, but the devil is in the details, even for the much simpler problem of FIB transport. The correct expressions are apparently not available in the literature (and incorrect expressions or approximations have been floated), justifying our derivations, at the risk of being seen as a pedagogical exercise of standard radiative-transfer theory. In the same vein we also show explicitly the transition between a volume integral and a quasi-thermal surface integral in the strong-trapping limit. We believe that deriving these results from first principles, starting with the Boltzmann collision equation, is an instructive exercise that offers many interesting insights that may be useful for future studies of astrophysical particle bounds.

\section{Radiative transfer by feebly interacting bosons}

We begin with the Boltzmann collision equation (BCE) for new bosons $a$ (reminiscent of ``axion'') that can be produced, for example, by processes of the type $\gamma+B\to B+a$, that is to say axion-photon conversion by interaction with fermions (for example semi-Compton scattering on electrons or muons) or other charged particles as in the Primakoff case, but photon coalescence $2\gamma\to a$ is also conceivable. On the other hand, scattering of the type $a+B\to B+a$ plays no role because the interaction is much more feeble than that of photons.

\subsection{Freeze out from first principles}

Ignoring FIB scattering from one momentum mode to another, we can focus on the evolution of a single mode with energy $\omega$ along some ray with spatial coordinate $x$. The BCE for the occupation number $f$ is in this case 
\begin{equation}\label{eq:Boltzmann}
    (\partial_t+v\partial_x)\,f=\Gamma_{\rm E}(1+f)-\Gamma_{\rm A} f=
    \Gamma_{\rm E}-\underbrace{(\Gamma_{\rm A}-\Gamma_{\rm E})}_{\hbox{$\Gamma_{\rm A}^*$}}  f,
\end{equation}
where $v$ is the particle velocity. Here $\Gamma_{\rm E}$ is the spontaneous emission rate that appears multiplied with the boson stimulation factor $1+f$, whereas $\Gamma_{\rm A}$ is the  absorption rate, and in general both depend on $\omega$ and $x$. In the second expression, the terms proportional to $f$ were consolidated and are proportional to the ``reduced absorption rate'' $\Gamma_{\rm A}^*=\Gamma_{\rm A}-\Gamma_{\rm E}$ that includes the effect of stimulated emission as a negative absorption rate. 

If the medium is in local thermal equilibrium, detailed balance implies that locally 
$\Gamma_{\rm E}=e^{-\omega/T}\Gamma_{\rm A}$ so that the reduced absorption rate is
\begin{equation}
    \Gamma\equiv\Gamma_{\rm A}^*=\Gamma_{\rm A}(1-e^{-\omega/T}),
\end{equation}
which we use as \textit{the\/} absorption rate and which is the quantity that defines the optical depth. The spontaneous emission rate is then expressed as
\begin{equation}\label{eq:GammaE}
    \Gamma_{\rm E}=\frac{\Gamma}{e^{\omega/T}-1},
\end{equation}
a relation between emission and absorption corresponding to Kirchhoff's Law.

In a stationary and homogeneous situation, the left-hand side (LHS) of Eq.~\eqref{eq:Boltzmann} vanishes and the equation is solved by a thermal Bose-Einstein distribution $\feq=(e^{\omega/T}-1)^{-1}$. So we may write the BCE instead for the deviation from equilibrium $\Delta f=f-\feq$ in the form
\begin{equation}\label{eq:Boltzmann-2}
    (\partial_t+v\partial_x)\,\Delta f=-\Gamma\, \Delta f.
\end{equation}
So it is the reduced absorption rate $\Gamma$ which damps the deviation of $f$ from equilibrium, explaining its central importance for radiative transfer.

In the context of thermal field theory, the boson propagation properties are encoded in its self-energy $\Pi$ within the medium. The imaginary part provides the rate-of-approach to thermal equilibrium as ${\rm Im}\,\Pi=-\omega \Gamma$ \cite{Weldon:1983jn}, once more highlighting the role of the reduced absorption rate as the central interaction parameter.

\subsection{Stationary state}
\label{sec:StationaryState}

Our main interest is a stationary situation, so only the gradient term on the LHS of the BCE survives and we need to solve
\begin{equation}
  v\frac{df}{dx}= \Gamma_{\rm E}-\Gamma f,
\end{equation}
where the spontaneous emission rate $\Gamma_{\rm E}$ is given in Eq.~\eqref{eq:GammaE} in terms of the reduced absorption rate $\Gamma$ under the assumption of local thermal equilibrium. 

To solve this equation we notice that $T$ and $\Gamma$ are functions of $x$ and we define the optical depth as
\begin{equation}
  \tau(x)=\int_{x}^{\infty}\frac{dx'}{\lambda(x')}
  \quad\hbox{with}\quad
   \lambda(x)=\frac{v}{\Gamma(x)},
\end{equation}
where $\lambda$ is the mean free path (MFP) for a FIB with velocity $v$. For massless FIBs we should use $v=c=1$ everywhere. The optical depth $\tau(x)$ is measured relative to a distant observer at $x=+\infty$ where $\tau(\infty)=0$. So finally one finds the solution
\begin{equation}\label{eq:final-occ}
  f(x)=\int_{-\infty}^{x}dx'\,\frac{\Gamma_{\rm E}(x')}{v}\,e^{\tau(x)-\tau(x')}.
\end{equation}
This is the intuitive answer that the occupation number at $x$ is filled by spontaneous production up to this point, reduced by the absorption along the path from production to detection. Here it was assumed that no radiation enters at the boundary at $x=-\infty$, i.e., all radiation is generated by emission within the realm of integration.

Instead of $x$ we may use $\tau(x)$ itself as a coordinate along the beam. Notice that this is a monotonically decreasing function of $x$ and thus uniquely invertible to provide $x(\tau)$. The limiting values are $\tau(\infty)=0$ and reaches a maximum value $\tau_{\rm max}=\tau(-\infty)$. Notice also that $d\tau(x)/dx=-\Gamma(x)/v=-1/\lambda(x)$ and we introduce the blackbody occupation number at $\tau$ for the local temperature $T(\tau)$
\begin{equation}
  \feq(\tau)=\frac{1}{e^{\omega/T({\tau})}-1}.
\end{equation}
We see that the solution
\begin{equation}\label{eq:fplus}
  f(\tau)=\int_{\tau}^{\tau_{\rm max}}d\tau'\,e^{\tau-\tau'}\,\feq(\tau')
\end{equation}
depends only on the temperature profile $T(\tau)$ along the ray. The velocity $v$ no longer appears explicitly because the optical depth is based on $\lambda$ and not on $\Gamma$. If the medium is very opaque so that we cannot see through the star to the other side we may use $\tau_{\rm max}=\infty$. For the occupation number at spatial infinity, corresponding to $\tau=0$, one finds in this opaque limit
\begin{equation}\label{eq:f0}
  f(0)=\int_{0}^{\infty}d\tau\,e^{-\tau}\,\feq(\tau).
\end{equation}
In the special case when the medium has the same $T$ everywhere this is simply $f(0)=\feq$, the Bose-Einstein occupation number. So an optically thick object at temperature $T$ radiates bosons with a thermal Bose-Einstein distribution. However, even if the radiating body has a hard material surface, the photons do not emerge from that surface but from a layer with thickness of a few MFPs.

We also consider the occupation number $f_-(\tau)$ of the ``backward mode'' moving in the opposite direction, toward the star,
\begin{equation}\label{eq:fminus}
  f_-(\tau)=\int_{0}^{\tau}d\tau'\,e^{-\tau+\tau'}\,\feq(\tau'),
\end{equation}
where it was assumed that at spatial infinity ($\tau=0$) the backward mode is not occupied. In this notation, the occupation
$f(\tau)$ of the outgoing mode is termed $f_+(\tau)$.

In this discussion we have implicitly assumed that the FIB absorption rate $\Gamma$ depends on the background medium which is geometrically bounded so that it makes sense to use a distant observer as a point of reference when using the optical depth as a measure of distance. However, when FIB decay of the form $a\to2\gamma$ is important, this approach is not justified. We will return to this question in the context of our spherically symmetric solution.

\subsection{Particle flux}

The radiation emerging from a source is usually not described in terms of the occupation numbers of the modes of the radiation field but rather by the corresponding particle or energy flux. The net particle flux in the outgoing direction is
\begin{equation}
    \phi=v\,(f_+-f_-)
\end{equation}
whereas the energy flux sports an additional factor $\omega$. Assuming no backward occupation at spatial infinity, the outgoing flux for a distant observer is simply $\phi(0)=v f(0)$ given in Eq.~\eqref{eq:f0}. At intermediate positions, the flux can be expressed as
\begin{equation}\label{eq:convolution}
    \phi(\tau)=v\,\int_{0}^{\infty}d\tau'\,{\rm sign}(\tau'-\tau)\,e^{-|\tau'-\tau|}\,\feq(\tau').
\end{equation}
So we find the intuitive result that the flux along some ray is driven by the temperature profile a few MFPs up- and downstream from the point of interest. Formally the function $\phi(\tau)$ on the interval $0\leq\tau<\infty$ is a certain linear transformation of the function $\feq(\tau)$ on that same interval.

\subsection{Example with power-law profile}

\begin{figure}[h]
\hbox to\textwidth{\hfil\includegraphics[height=0.3\textwidth]{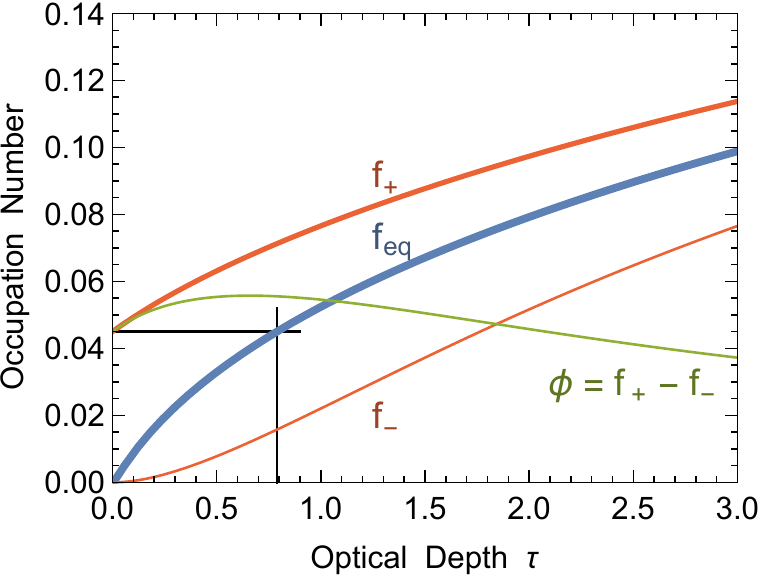}
\hskip12pt\includegraphics[height=0.3\textwidth]{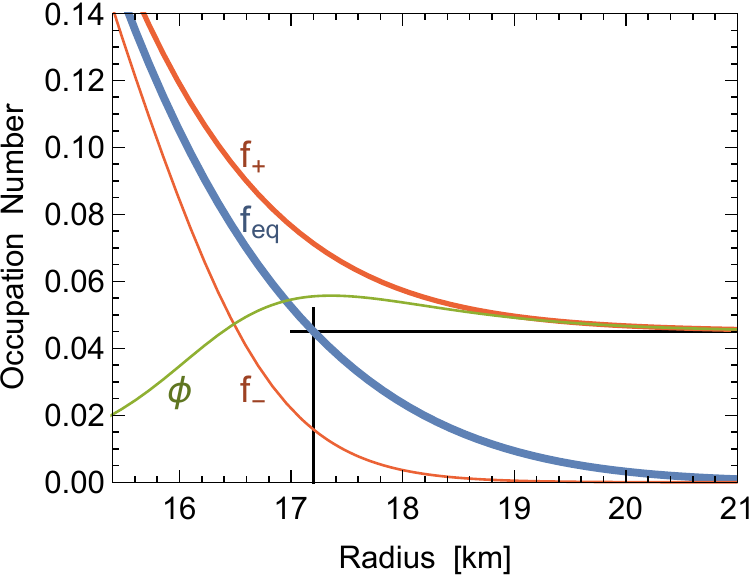}\hfil}
  \caption{Solutions for the occupation numbers $f_\pm$ and the flux $\phi$ for a massless boson and using our power-law profile for the temperature and using a typical energy $\omega=3T_1$. The horizontal black line projects the escaping flux (the occupation number at the stellar surface) with the equilibrium one and marks the Stefan-Boltzmann optical depth for this energy, here approximately at $\tau\simeq0.8$}\label{fig:f-solution}
\end{figure}

We can illustrate FIB freeze out with a $T$ profile inspired by a realistic Proto Neutron Star (PNS) profile of the form
\begin{equation}\label{eq:PowerLaw}
  T(\tau)=T_1\tau^p,
\end{equation}
where $0<p\ll 1$ is a small number for which we use $p=1/5$ and $T_1$ is the temperature at unit optical depth. Moreover, we assume the FIB to be massless so that $v=1$. For a typical boson energy of $\omega=3T_1$ we show the solutions $f_\pm$ as well as the flux $\phi=f_+-f_-$ in Fig.~\ref{fig:f-solution}. We see that the flux escaping from the star corresponds approximately to $\feq$ at $\tau\simeq0.8$, but at this location the actual solution is far away from this value. The approach to the asymptotic solution is slow in the decoupling region.

For illustration we can also go back to coordinate space and show these results as a function of geometric radius. We find it useful to take inspiration from a realistic model of a SN core, following in particular the Garching group's muonic model SFHo-18.8 \cite{Bollig:2020xdr,CCSNarchive} that we employed earlier for other studies \cite{Caputo:2021rux, Caputo:2022mah}. In this case, one can see that the temperature varies approximately as $T=T_1\,(r_1/r)^4$ which is equivalent to $T=T_1 \tau^{1/5}$, implying $\tau=(r_1/r)^{20}$ where we take $r_1=17$~km. In this representation, the approach to the asymptotic solution looks more intuitive, but it remains true that the approach to the asymptotic solution does not happen at the nominal decoupling radius, but is considerably smeared out even though here we have a fixed energy and no energy dependence of the cross section.

So the picture that a Stefan-Boltzmann flux emerges from some narrow geometric range like ``surface emission'' is clearly not accurate. The bosons reaching infinity derive from a broad radial range, equivalent to a broad range of optical depth.

\section{Strong trapping regime and plane-parallel atmosphere}
\label{sec:PlaneParallel}

The single-ray solutions of the previous section provide the full answer to the question of the stationary FIB radiation field based on a source distribution with prescribed properties (no feedback effects by particle emission on the medium). It remains to cast this result into a more explicit form for relevant overall geometries. To discuss more explicitly radiation decoupling in the strong trapping limit we use a plane-parallel atmosphere, where the temperature and optical depth are only functions of a cartesian coordinate $z$ perpendicular to the atmospheric layering. In the example shown in Fig.~\ref{fig:f-solution}, inspired by a realistic SN core model, the decoupling radius is some 17~km and the relevant shell has a thickness of a few km, so the plane-parallel approximation should provide a reasonable first description.

\subsection{Intensities vs.\ occupation numbers}

Solving the Boltzmann collision equation was most transparent using occupation numbers which appear directly in Bose stimulation factors. However, in the end we ask for the energy flux at some radial position. In this spirit we turn from occupation numbers to radiation intensities for a mode $\bk$ of the radiation field
\begin{equation}
    I_\bk=4\pi\,\frac{\omega^2|\bk|}{(2\pi)^3}\,f_\bk,
\end{equation}
where $\omega=(m_a^2+\bk^2)^{1/2}$. Notice that $|\bk|=v\omega$ where $v$ is the boson velocity. We have normalized the intensity such that the integral over energy and directions $\int I_\bk\, d\omega\,d\Omega/4\pi$ is the local energy density. Whether or not to include the factor of $4\pi$ in the definitions of $I_\bk$ and the blackbody intensity $B_\omega$ in Eq.~\eqref{eq:B-source} is a matter of convenience. 

When the FIBs are in thermal equilibrium, the occupation number is $f_\bk=1/(e^{\omega/T}-1)$ and the equilibrium (blackbody) intensity is denoted as
\begin{equation}\label{eq:B-source}
    B_\omega^v=\frac{\omega^2\sqrt{\omega^2-m_a^2}}{2\pi^2}\,\frac{1}{e^{\omega/T}-1}=v_\omega B_\omega,
    \quad\hbox{where}\quad
    B_\omega=\frac{\omega^3}{2\pi^2}\,\frac{1}{e^{\omega/T}-1}.
\end{equation}
Here $B_\omega$ is the blackbody intensity for one massless degree of freedom and $v_\omega=\sqrt{1-m_a^2/\omega^2}$ is the velocity for a boson with mass $m_a$. For the massless case, the total energy density~is
\begin{equation}\label{eq:B0}
   B=\int_0^\infty \!d\omega\,B_\omega=\frac{\pi^2}{30}\,T^4.
\end{equation}
For a nonvanishing mass, no simple expression exists.

\subsection{Angular moments}
\label{SubSec:Angular}

The previous single-ray solution applies to a mode propagating in the radial direction, but now we consider one that is inclined by $\mu=\cos\theta$ such that $\mu=+1$ is the outward direction and $\mu=-1$ the inward one. We begin with Eq.~\eqref{eq:final-occ} for the occupation at position $x=z/\cos\theta$ along the ray. As variable of integration we may use $z$, so we use the vertical depth as a measure of propagation distance, or equivalently, the optical depth $\tau$ in the vertical direction. Following the previous steps we find for the outgoing and incoming intensities
\begin{equation}\label{eq:final-occ-4}
  I^+_{\omega,\mu}(\tau)=\frac{1}{\mu}\int_{\tau}^{\infty}d\tau'\,e^{(\tau-\tau')/\mu} B^v_\omega(\tau')
\quad\text{and}\quad
  I^-_{\omega,\mu}(\tau)=\frac{1}{\mu}\int_{0}^{\tau}d\tau'\,e^{-(\tau-\tau')/\mu} B^v_\omega(\tau'),
\end{equation}
where $\mu>0$ by definition, i.e., we treat inward-moving modes explicitly as backward moving ones with positive $\mu$. 

We are mostly interested in the energy flux, but in general one defines angular moments of the type
\begin{equation}\label{eq:moments}
   M_\omega^{(n)}=\frac{1}{2}\,\int_{-1}^{+1}d\mu\,(v_\omega\mu)^n\,I_{\omega,\mu}
   =\frac{1}{2}\,\int_{0}^{1}d\mu\,(v_\omega\mu)^n\Bigl[I^+_{\omega,\mu}+(-1)^n I^-_{\omega,\mu} \Bigr].
\end{equation}
Traditionally the zeroth moment (the energy density) is called $J_\omega$, the first moment (the energy flux) $H_\omega$, and the second moment $K_\omega$ is related to pressure. For photons $v=1$ and $B_\omega$ acquires a factor of 2 for the two polarization states. The factors of $v$ are understood in the sense that a flux (of energy or particles) requires one power of $v$ compared with the massless case, whereas the pressure, being essentially a flux of momenta, requires one more $v$ factor. Indeed, the spatial part of the stress-energy tensor dimensionally involves (momentum)$^2$.

\begin{figure}[ht]
\vskip12pt\vskip12pt
\centering
\includegraphics[width=0.45\textwidth]{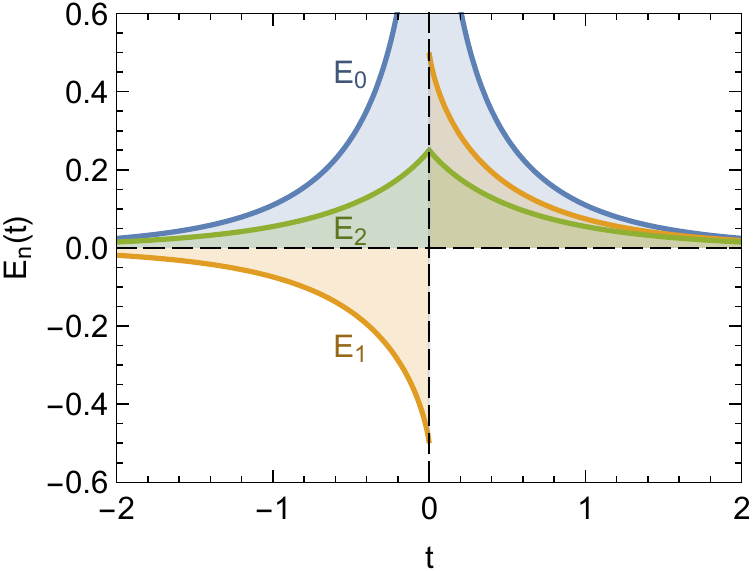}
  \caption{The $n^{\rm th}$-order integral kernels $\sE_n(s)$ defined in Eq.~\eqref{eq:kernels}.}\label{fig:kernels}
\end{figure}

The angle integrations in Eq.~\eqref{eq:final-occ-4} can be performed explicitly. For the $n^{\rm th}$  moment and using $w=1/\mu$ one finds\
\begin{equation}\label{eq:expint}
    \int_0^1 d\mu\,\mu^n\,\frac{e^{-t/\mu}}{\mu}=\int_1^\infty dw\,\frac{e^{-t w}}{w^{n+1}}=E_{n+1}(t),
\end{equation}
where $E_m(t)$ is the $m^{\rm th}$ exponential integral, in {\sc Mathematica} notation {\tt ExpIntegralE[m,t]}. It obeys $dE_m(t)/dt=-E_{m-1}(t)$ and $E_m(t)=[e^{-t}-t E_{m-1}(t)]/(m-1)$ for \hbox{$m>1$}. We use $E_m(t)$ only for positive arguments where it is positive and real. To consolidate the $\pm$ cases in Eq.~\eqref{eq:moments} in a single expression it is convenient to define integral kernels of the form
\begin{equation}\label{eq:kernels}
    \sE_n(t)=\frac{1}{2}\,{\rm sign}(t)^n\,E_{n+1}(|t|)
    \quad\hbox{where}\quad
    {\rm sign}(t)=\frac{t}{|t|},
    \end{equation}
shown in Fig.~\ref{fig:kernels}. These are even functions of $t$ for even $n$ and odd functions of $t$ for odd $n$. With this notation, the moments of Eq.~\eqref{eq:moments} are
\begin{equation}\label{eq:moments-1}
   M_\omega^{(n)}(\tau)=v_\omega^{n+1}\int_{0}^{\infty}d\tau'\,\sE_{n}(\tau'-\tau)\, B_\omega(\tau').
\end{equation}
Notice that one factor of $v_\omega$ comes from $B_\omega^v$ for particles with mass, whereas $B_\omega$ is the massless intensity and thus only a property of the medium profile, not the particle mass. In the massless case, 
these are the Schwarzschild-Milne equations, providing us with the moments of the radiation field as linear transformations of the blackbody intensity on the interval $0\leq\tau<\infty$. The $0^{\rm th}$-order case, providing the local energy density, is known as the $\Lambda$-transformation.

\clearpage

\subsection{Diffusion regime}

Asymptotically  $E_n(t)= t^{-1}e^{-t}$ for $t\to\infty$ independently of $n$. Among other consequences, this implies that integrals over any power $t^n$ weighted with such kernels converge. It also implies that the local values of the moments only depend on the thermal radiation field a few MFPs up- and downstream. In particular, we consider a general function $b(t)$ that we expand as a Taylor series
\begin{equation}
  b(t)=\sum_{m=0}^{\infty}\,\frac{b^{(m)}(0)\,t^m}{m!}.
\end{equation}
Then we find 

\begin{eqnarray}
\int_{-\infty}^{+\infty}dt\,b(t)\,\sE_n(t)&=& \sum_{m=0}^{\infty}\,\frac{1+(-1)^{m+n}}{2}\,\frac{b^{(m)}(0)}{m+n+1},
\end{eqnarray}

which for the first three momenta gives explicitly
\begin{subequations}
\begin{eqnarray}
\int_{-\infty}^{+\infty}dt\,b(t)\,\sE_0(t)&=& \sum_{m=0}^{\infty}\,\frac{1+(-1)^m}{2}\,\frac{b^{(m)}(0)}{m+1}=
b(0)\,\,+\frac{b''(0)}{3}\,\,+\ldots\\
\int_{-\infty}^{+\infty}dt\,b(t)\,\sE_1(t)&=& \sum_{m=0}^{\infty}\,\frac{1-(-1)^m}{2}\,\frac{b^{(m)}(0)}{m+2}=
\frac{b'(0)}{3}+\frac{b'''(0)}{5}+\ldots\\
\int_{-\infty}^{+\infty}dt\,b(t)\,\sE_2(t)&=& \sum_{m=0}^{\infty}\,\frac{1+(-1)^m}{2}\,\frac{b^{(m)}(0)}{m+3}=
\frac{b(0)}{3}\,\,+\frac{b''(0)}{5}\,+\ldots
\end{eqnarray}
\end{subequations}

Of course, this representation makes only sense at large optical depth where the lower limit of integration can be extended to $-\infty$ and the Taylor expansion is really around a point $\tau\gg 1$. In this case we see that the kernels for the first two momenta at leading order effectively act as
\begin{equation}
    \sE_0(\tau)\simeq\delta(\tau),\quad  
    \sE_1(\tau)\simeq-{\textstyle\frac{1}{3}}\delta'(\tau),
\end{equation}
assuming the function $b(\tau)$ varies sufficiently slowly. Recall that $\int dx\,f(x)\,\delta'(x)=-f'(x)$.

So deeply in the trapped regime, many MFPs away from the surface, the net diffusive flux is
\begin{equation}
F_{\rm diff}(\tau,\omega)=\frac{v^2_\omega}{3}\,\frac{d}{d\tau}\,B_\omega(\tau)
\quad\hbox{or}\quad
F_{\rm diff}(z,\omega)=-\frac{v_\omega^2\lambda_\omega}{3}\,\frac{d}{dz}\,B_\omega(z)
\end{equation}
which is driven by the temperature gradient. (We prefer the letter $F$ to $H$ that is traditional in the theory of radiative transfer.) Recall that $z$ is the coordinate perpendicular to the plane-parallel atmospheric layering, that the MFP is $\lambda_\omega=v_\omega/\Gamma_\omega$ with the particle velocity $v_\omega$, that the Jacobian is $dz/d\tau=-\lambda$, and that a factor $v_\omega^2$ comes from the first factor in Eq.~\eqref{eq:moments-1}. The diffusive flux is a good representation of the true flux when the MFP is short compared with the length scale of temperature variation. However, we can formally define $F_{\rm diff}(\tau)$ everywhere, whether or not it is a good approximation of the true flux.

Finally we can define the Stefan-Boltzmann (SB) flux, given by the equilibrium intensity at a given radial position times an average angular flux factor 1/2 and times another factor 1/2 to count only the outward going modes. $F_{\rm SB}(\tau)$ is the hypothetical FIB flux produced by a black surface at the radial position $\tau$ with the local $T(\tau)$. Of course, the SB flux is simply another way of expressing the local temperature. Overall we define three different fluxes
\begin{subequations}
\begin{eqnarray}
  F_{\rm true}(\tau,\omega) &=& v_\omega^2\int_{0}^{\infty}d\tau'\,\sE_1(\tau'-\tau) B_\omega(\tau'),
  \label{eq:Ftrue-convolution}
  \\[1.5ex]
  F_{\rm diff}(\tau,\omega) &=&\frac{v_\omega^2}{3}\,\frac{d}{d\tau}\,B_\omega(\tau),
  \label{eq:Fdiff-1}\\[1.5ex]
  F_{\rm SB}(\tau,\omega)   &=&\frac{v_\omega^2}{4}\,B_\omega(\tau).
\end{eqnarray}
\end{subequations}
For the diffusive flux, we have rediscovered the usual factor 1/3 following directly from the properties of the exponential integrals. We recall that $B_\omega$ is the intensity for one massless boson degree of freedom. 

\subsection{Integration over energies for a grey atmosphere}

We are usually not interested in the detailed energy dependence unless there are resonant effects. So we may integrate over energies, but this requires to specify the energy dependence of the FIB interaction rate. The assumption that the reduced absorption rate $\Gamma$ does not depend on energy is called the ``grey-atmosphere approximation'' in the theory of radiative transfer. Moreover, we now consider massless particles with $v=c=1$. The grey-atmosphere approximation is surprisingly well motivated by FIBs absorbed by the Primakoff process as detailed in Sec.~\ref{sec:PrimakoffInteractionModel}. Here we simply use this approach for the purpose of illustration.

The integrated blackbody energy density for a single massless boson degree of freedom was given in 
Eq.~\eqref{eq:B0} as $B(\tau)=(\pi^2/30)\,T(\tau)^4$, where the optical depth does not depend on energy by assumption. Then our three fluxes are explicitly
\begin{subequations}\label{eq:three-fluxes}
\begin{eqnarray}
  F_{\rm true}(\tau) &=& \int_{0}^{\infty}d\tau'\,\sE_1(\tau'-\tau)\,B(\tau'),
  \label{eq:three-fluxes-true}\\[1.5ex]
  F_{\rm diff}(\tau) &=& \frac{1}{3}\,\frac{d}{d\tau}\,B(\tau),\\[1.5ex]
  F_{\rm SB}(\tau)   &=&\frac{1}{4}\,B(\tau).
\end{eqnarray}
\end{subequations}
Besides overall coefficients, the SB flux is a purely local quantity, the diffusive flux a spatial derivative, whereas the true flux involves a nonlocal operator, a convolution over all space, in practice a few units of optical depth upstream and downstream. So these three fluxes are nicely systematic about the FIB flux in the trapping limit.

For a distant observer ($\tau=0$) and inserting the definition of ${\sf E}_1(t)$, the true flux is found to be
\begin{equation}\label{eq:Ftrue-distant}
    F_{\rm true}(0)=\frac{1}{2}\int_{0}^{\infty}d\tau\,E_2(\tau)\,B(\tau),
\end{equation}
where $E_2(t)$ is the second exponential integral. The interpretation is that of every boson launched isotropically at optical depth $\tau$, the probability to escape to infinity is given by the transmittance ${\sf T}(\tau)=\frac{1}{2}\,E_2(\tau)$. The factor 1/2 accounts for all bosons launched away from the surface cannot escape, whereas the others have a chance of escape of $E_2(\tau)$. If all bosons were emitted either exactly toward or exactly away from the surface, the transmittance would be $\frac{1}{2}\,e^{-\tau}$. Due to the angular average $e^{-\tau}\to E_2(\tau)$. We recall that $\tau$ here means the optical depth counted directly inward from the surface. The functional form of $\frac{1}{2}\,E_2(\tau)$ is, for positive~$\tau$, the orange curve in Fig.~\ref{fig:kernels} marked ${\sf E}_1$. For large arguments it is  $\frac{1}{2}e^{-\tau}/\tau$.

\subsection{Example with power-law profile}

\begin{figure}[ht]
\centering
\hbox to\textwidth{\hfil\includegraphics[height=0.3\textwidth]{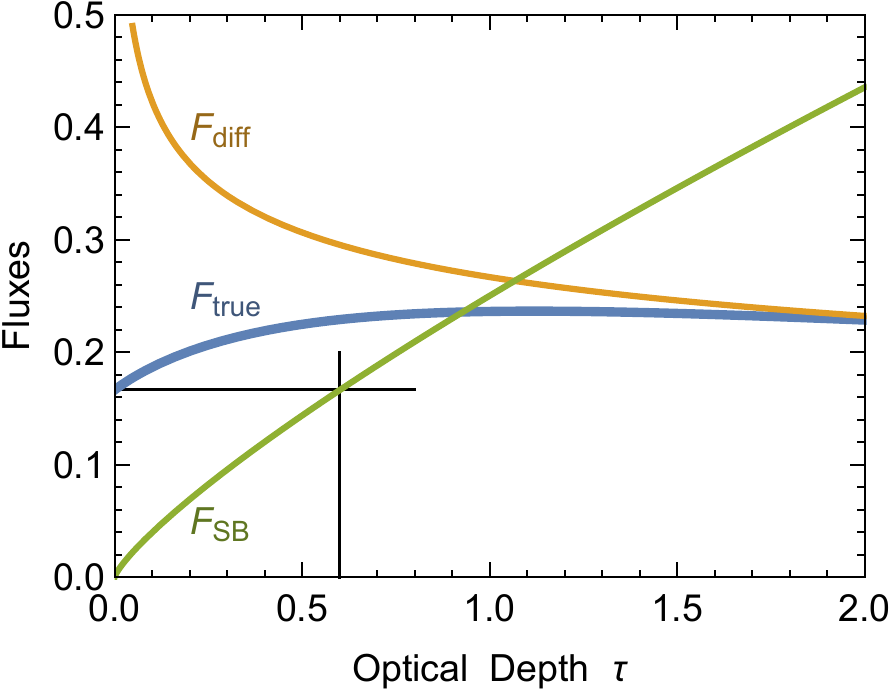}
\hskip6pt\includegraphics[height=0.3\textwidth]{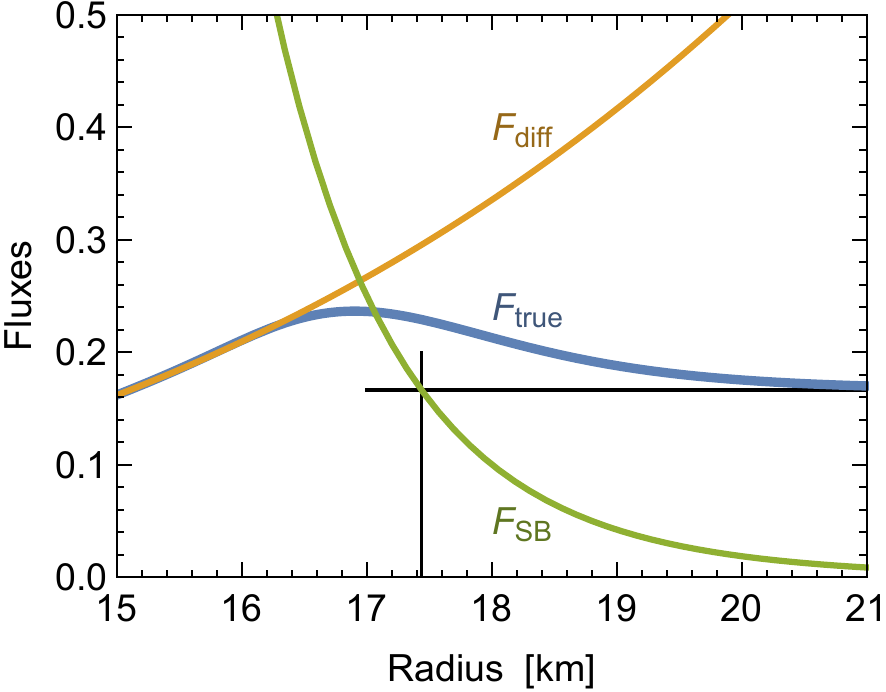}\hfil}
  \caption{The fluxes of Eq.~\eqref{eq:fluxes4p} for $p=0.2$. The optical depth where the SB flux matches the escaping true flux is $\tau_{\rm SB}=0.60$. For the radial dependence we used $\tau=(17~{\rm km}/r)^{20}$ as earlier.}\label{fig:Fluxes4p}
\end{figure}

For illustration we return to the power-law profile of Eq.~\eqref{eq:PowerLaw}. Apart from a global factor that we now leave out, the three fluxes are
\begin{equation}\label{eq:fluxes4p}
  F_{\rm true}(\tau,p)=\int_{0}^{\infty}d\tau'\,\sE_1(\tau'-\tau)\,\tau'^{\,4p},\quad
  F_{\rm diff}(\tau,p)=\frac{4p}{3}\,\tau^{4p-1},\quad
  F_{\rm SB}(\tau,p) =\frac{1}{4}\,\tau^{4p}.
\end{equation}
For a typical case $p=0.2$ we show the fluxes as a function of optical depth and radius in Fig.~\ref{fig:Fluxes4p}. We see that the diffusive and true fluxes become asymptotically close at large optical depth and then separate in the freeze-out region. This is most intuitively clear in the radial plot. The required optical depth for the SB flux to match the escaping true flux is $\tau_{\rm SB}\simeq0.60$.

Notice that $p=1/4$ is a special value where $F_{\rm diff}=1/3$ is a constant and $F_{\rm SB}=\tau/4$ increases linearly. We have not used this value to avoid an overly special case. In general, the true flux at the surface ($\tau=0$) is explicitly
\begin{equation}\label{eq:tautau}
  F_{\rm true}(0,p)=\frac{p\,\Gamma(4p)}{1+2p}\simeq
  \frac{1}{6}+0.06(p-1/4)+1.02(p-1/4)^2+{\cal O}[(p-1/4)^3],
\end{equation}
where we have used an expansion around the special value of $p=1/4$ where this flux is near to a minimum.
The condition $F_{\rm SB}(\tau_{\rm SB})=\frac{1}{4}\tau_{\rm SB}^{4p}=F_{\rm true}(0,p)$ leads to
\begin{equation}
  \tau_{\rm SB}(p)=
  \left(\frac{4p\,\Gamma(4p)}{1+2p}\right)^{1/4p}\simeq
    \frac{2}{3}\,\left[1+2\left(p-\frac{1}{4}\right)\right]
\end{equation}
where the approximation is good on the few-percent level in the entire range $0<p<1$. For our special value $\tau_{\rm SB}(1/4)=2/3$ is exact.

In the neutrino decoupling region of a SN core, when diffusive transport is still appropriate, the neutrino flux itself, driven by the temperature gradient, is approximately constant. Therefore, the radiation density of neutrinos scales roughly linearly with neutrino optical depth. As the neutrino scattering rate is proportional to the density as assumed for our FIBs, the power-law index $p\simeq1/4$ for the temperature as a function of optical depth is well motivated in the neutrino decoupling region and borne out from numerical models.

\begin{figure}[ht]
\centering
\hbox to\textwidth{\hfil\includegraphics[height=0.3\textwidth]{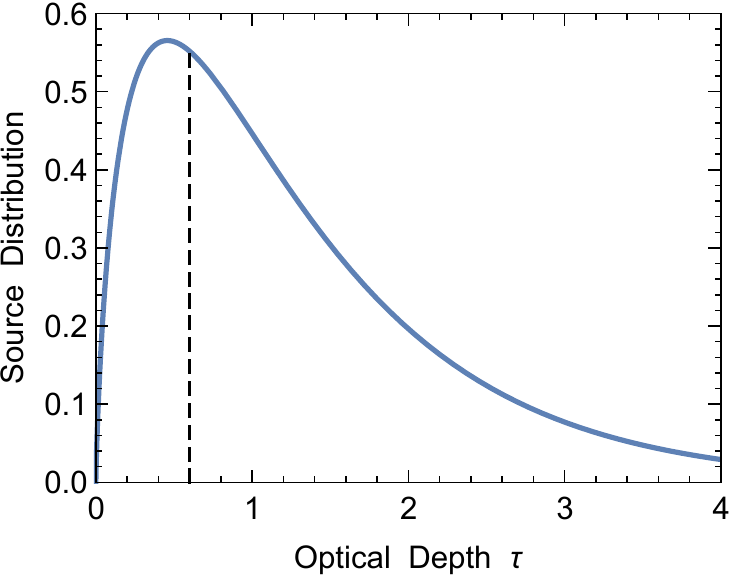}
\hskip12pt\includegraphics[height=0.3\textwidth]{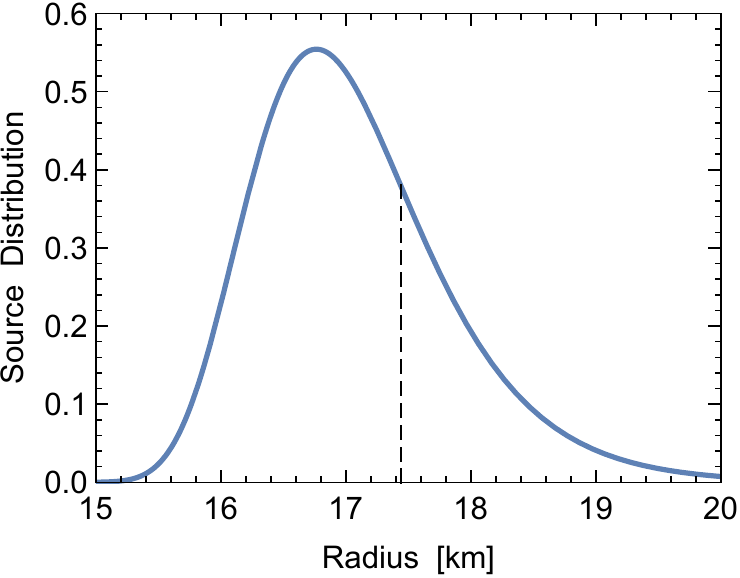}\hfil}
  \caption{Source distribution of bosons reaching a distant observer. For the temperature distribution, the power-law index $p=0.2$ was used as in Fig.~\ref{fig:Fluxes4p}. For the dependence on the geometric radius we used again $\tau=(17~{\rm km}/r)^{20}$. The vertical dashed lines indicate the position of the Stefan-Boltzmann radius of $\tau_{\rm SB}\simeq0.60$ and $r_{\rm SB}\simeq17.43\,~{\rm km}.$}\label{fig:SourceDistribution}
\end{figure}

We may also ask where the emitted radiation reaching a distant observer is actually emitted. Equation~\eqref{eq:Ftrue-distant} implies a distribution proportional to $E_2(\tau)B(\tau)$. For $p=1/5$ and thus $B\propto\tau^{4/5}$, the normalized source distribution is
\begin{equation}
    f_{\rm source}(\tau)=\frac{7}{2\,\Gamma(4/5)} \tau^{4/5} E_2(\tau),
\end{equation}
shown in the left panel of Fig.~\ref{fig:SourceDistribution}. As a function of geometric radius once more we assume $\tau=(r_0/r)^{20}$ with $r_0=17\,{\rm km}$, leading to the normalized source distribution
\begin{equation}
    f_{\rm source}(r)=\frac{70}{\Gamma(4/5)}\,\frac{1}{r_0}  \left(\frac{r_0}{r}\right)^{37}
        E_2\bigl[(r_0/r)^{20}\big]
\end{equation}
shown in the right panel of  Fig.~\ref{fig:SourceDistribution}. The vertical dashed lines show the location of the Stefan-Boltzmann radius.

We learn from this figure that the bosons reaching infinity originate from a fairly thick shell, not a sharp ``boson sphere.'' The grey-atmosphere model, without any energy dependence of the cross section, provides the ``sharpest'' conceivable emission sphere. For neutrinos, the cross section varies with the square of energy and the ``neutrino sphere'' is much more smeared out and energy dependent.

\subsection{Constant plus linear profile for the radiation density}

The special power-law profile $T(\tau)\propto \tau^{1/4}$ corresponds to a linear profile for the radiation density $B(\tau)\propto\tau$. The next simple profile derives from adding an arbitrary constant
\begin{equation}\label{eq:linear-profile}
  B(\tau)=B_0\,\left(\tau+q\right),
\end{equation}
where the letter $q$ is traditionally used. The true flux is found through the convolution of Eq.~\eqref{eq:three-fluxes-true}, leading to a complicated expression in terms of exponential integrals. At the surface ($\tau=0$) one finds the following true flux to be compared with the SB flux
\begin{equation}\label{eq:SB-linear}
  F_{\rm true}(0)=B_0\,\left(\frac{1}{6}+\frac{q}{4}\right)
  \quad\hbox{while}\quad
  F_{\rm SB}=B_0\,\left(\frac{\tau_{\rm SB}}{4}+\frac{q}{4}\right).
\end{equation}
Thus the true flux at the surface is the same as the SB flux at the optical depth $\tau_{\rm SB}=2/3$,
independently of the constant $q$. This is the formal derivation of where this particular reference number comes from that floats around in the literature. For other temperature profiles and for non-grey atmospheres, $\tau_{\rm SB}=2/3$ is only an estimate.

\subsection{Self-consistent temperature profile and Eddington case}

In this paper we are considering FIB emission from a star or SN core with prescribed properties. On the other hand, in the trapping regime the FIB transfer of energy is not a perturbative effect, especially when they decouple at a radius larger than the neutrino sphere. In this case, the atmospheric run of temperature is determined by FIB energy transport and, in a stationary state, is determined by the condition $F_{\rm true}(\tau)=\text{constant}$. Finding the corresponding $B(\tau)$ is a formidable mathematical challenge that was solved in different ways as detailed, for example, in the book \cite{Kourganoff:1952}. Expressing the solution in the form of Eq.~\eqref{eq:linear-profile}, the solution $q(\tau)$ is called the Hopf function that can be explicitly expressed, for example, as an integral that can be evaluated numerically.

\begin{figure}[ht]
\vskip12pt
\centering
\includegraphics[width=0.45\textwidth]{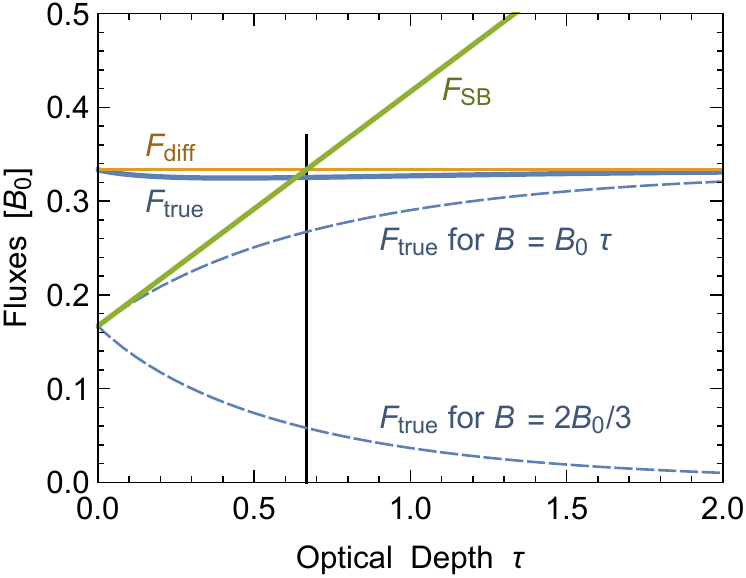}
  \caption{Fluxes for the Eddington profile $B(\tau)$ of the form Eq.~\eqref{eq:linear-profile} with $q=2/3$. The optical depth where the SB flux matches the escaping true flux is $\tau_{\rm SB}=2/3$ exactly. We show separately the fluxes generated by the linear and constant bits of the radiation density that each contribute exactly $B_0/6$ to the flux at the surface. If one were to use the Hopf function $q(\tau)$ instead of $q=2/3$, the true flux (solid blue line) would exactly equal the constant $B_0/3$ (orange horizontal line) that is also equal to the nominal $F_{\rm diff}$, which is here constant everywhere and shown even in the low-$\tau$ region where the diffusion approximation is not justified.}\label{fig:constantflux}
\end{figure}

We mention in passing that there is a surprisingly accurate approximation credited to  Milne and Eddington that is given by the constant $q=2/3$. From Eq.~\eqref{eq:SB-linear} we glean that in this case the true surface flux is $F_{\rm true}(0)=B_0/3$ and thus the same as the diffusive flux deep inside. The different flux components are shown in Fig.~\ref{fig:constantflux}, where the constant and linear terms of $B(\tau)$ each provide exactly the flux $B_0/6$ at the surface. While the Eddington profile was chosen to provide the same flux at the surface as deep inside, we see from Fig.~\ref{fig:constantflux} that the flux is surprisingly constant also in the intermediate range. We see that the SB flux, shown as a green line, matches the surface flux (horizontal orange line) at $\tau_{\rm SB}=2/3$ as expected.

\clearpage

\subsection{True-flux convolution in geometric variables}

Estimating the FIB flux with the SB approach is a good approximation that can be done for the energy-integrated flux or, if the monochromatic reduced absorption rate $\Gamma_\omega$ strongly depends on energy, for every $\omega$ separately. However, many of the recent papers that have motivated our study used a time series of numerical SN models that were post-processed to obtain the FIB luminosity in the trapping limit. So if one anyway performs a numerical study of that type, one may as well compute directly the true flux for each energy $\omega$ based on the convolution integral Eq.~\eqref{eq:Ftrue-convolution}. 

However, while the optical depth $\tau$ as a measure of distance is very useful for conceptual discussions, it is somewhat abstract for practical implementation. More importantly, it has the disadvantage that $\Gamma_\omega$ is assumed to decrease with increasing radius like the medium of a star so that spatial infinity corresponds to vanishing optical depth. However, for massive FIBs that can decay, for example by $a\to2\gamma$, the concept of optical depth is not directly appropriate and the FIB flux at infinity vanishes irrespective of the details of the source.

Both issues are resolved by returning to an integral over a geometric variable $z$ which here is the coordinate perpendicular to the plane-parallel atmosphere. The convolution integral of Eq.~\eqref{eq:Ftrue-convolution} for the monochromatic true flux becomes explicitly
\begin{eqnarray}\label{eq:Ftrue-geometric}
  F_\omega(z)&=&\int_{-\infty}^{+\infty}dz'\,\Gamma_\omega(z')\,v_\omega B_\omega(z')\,
  \sE_1\biggl[\int_{z'}^{z}\!\! dz''\,\frac{\Gamma_\omega(z'')}{v_\omega} \biggr]
  \nonumber\\[1.5ex]
  &=&\int_{-\infty}^{+\infty}dz'\,Q_\omega(z')\,
  \sE_1\biggl[\int_{z'}^{z}\frac{dz''}{\lambda_\omega(z'')} \biggr].
\end{eqnarray}
Here $\omega$ is the energy of a FIB with mass $m_a$ and velocity $v_\omega=(1-m_a^2/\omega^2)^{1/2}$. The reduced absorption rate $\Gamma_\omega(z)$ can also include free decay far away from the source. The thermal intensity $B_\omega(z)=(\pi^2/30)T(z)^4$ is the one for a massless boson. Notice that one velocity factor in front of Eq.~\eqref{eq:Ftrue-convolution} has cancelled against $v_\omega^{-1}$ appearing in the Jacobian through $|d\tau_\omega/dz|=1/\lambda_\omega=\Gamma_\omega/v_\omega$. Moreover, $\lambda_\omega(z)$ is the local MFP, based on the reduced absorption rate. 

We have also introduced the volume energy loss rate, differential with regard to its variable $\omega$,
\begin{equation}\label{eq:Qdefinition}
    Q_\omega(z)=\Gamma_\omega(z)\,v_\omega B_\omega(z),
\end{equation}
where the thermal FIB energy density was defined in Eq.~\eqref{eq:B-source}. Recall that the spontaneous emission rate is $\Gamma_{{\rm E},\omega}=\Gamma_\omega/(e^{\omega/T}-1)$, to be multiplied with the phase-space factor $v_\omega\omega^3/(2\pi^2)$ to obtain the energy emission rate per energy interval $d\omega$. Together this implies Eq.~\eqref{eq:Qdefinition} as a product of the reduced absorption rate times the blackbody FIB intensity. Notice that this applies to any process that absorbs the FIBs, including inverse bremsstrahlung or two-photon decay. The overall normalization (including a factor of $4\pi$ in $B_\omega$) is such that
\begin{equation}
    Q(z)=\int_{m_a}^\infty d\omega\, Q_\omega(z)
\end{equation}
is the local energy loss rate per unit volume, for example in units of ${\rm erg}\,{\rm cm}^{-3}\,{\rm s}^{-1}$.

\begin{figure}[ht]
\centering
\includegraphics[width=0.45\textwidth]{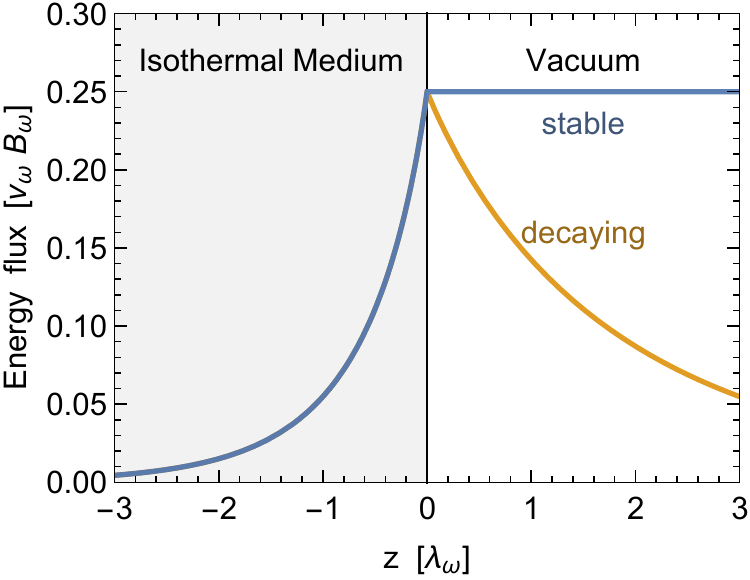}
  \caption{Uniform and isothermal medium with a surface at $z=0$. {\em Blue line:} Boson energy flux for a given MFP $\lambda_\omega$ in the medium and no interaction in vacuum. {\em Orange line:} Same MFP in the medium, but a remaining MFP of $3\lambda_\omega$ due to decays in vacuum.}\label{fig:isothermal}
\end{figure}

The non-appearance of a velocity factor in the flux expression of Eq.~\eqref{eq:Ftrue-geometric} is slightly confusing. Therefore, as a sanity check, we consider a uniform plane-parallel atmosphere at a constant temperature. The atmosphere ends at a surface at $z=0$. So $B_\omega(z)=0$ for $z>0$ and is constant for $z<0$. Likewise, $\Gamma_\omega$ is constant for $z<0$ and vanishes otherwise. The convolution integral can be solved analytically, however requiring many cases depending on the values of $z$, $z'$ and $z''$. We find explicitly
\begin{equation}\label{eq:Fisothermal}
    F_\omega(z)=\frac{v_\omega^2 B_\omega}{4}
    \begin{cases}2 E_3(-z/\lambda_\omega) &\hbox{for $z< 0$}, \\
    1&\hbox{for $z\geq 0$,}
    \end{cases}
\end{equation}
where $E_3$ is the third exponential integral discussed around Eq.~\eqref{eq:expint}. We show this solution in Fig.~\ref{fig:isothermal} where we see that $F_\omega(z)$ develops a few MFPs below the surface and emerges with the Stefan-Boltzmann value $v^2_\omega B_\omega/4$, including a factor $v^2_\omega$ in front of the massless-boson intensity. For a massive particle, the flux is reduced in two ways, the explicit $v_\omega$ coming from the flux and one factor from the phase-space density of modes within an energy interval $d\omega$, not from the velocities of individual particles.

\subsection{Including boson decay}

We briefly illustrate the case where the FIBs can decay after emerging from the surface of an isothermal medium. So we assume that in the medium the (reduced) MFP is $\lambda_\omega$, caused by all kinds of processes, including photon coalescence. In vacuum, only free decay is possible for which we take schematically an MFP of $3\lambda_\omega$. Performing the convolution, in analogy to 
Eq.~\eqref{eq:Fisothermal} we find
\begin{equation}\label{eq:Fisothermaldecay}
    F_\omega(z)=\frac{v_\omega^2 B_\omega}{4}
    \begin{cases}2 E_3(-z/\lambda_\omega) &\hbox{for $z< 0$} \\
    1&\hbox{for $z= 0$}\\
    2 E_3(z/3\lambda_\omega) &\hbox{for $z>0$}
    \end{cases}
\end{equation}
as shown in Fig.~\ref{fig:isothermal}. The behavior in the medium depends only on the reduced interaction rate, not the individual contributions from different processes. In vacuum, where the source $B_\omega=0$, only vacuum decay is relevant. Notice that the variation with distance is not exponential because the large-argument limit is $E_3(s)\to e^{-s}/s$. The particles still decay exponentially on their trajectories, but the angle average implies that the overall flux decreases faster with distance. This behavior is a consequence of the plane-parallel model because at a large distance from a star, many stellar radii away, the flux decreases exponentially because the trajectories become more and more collinear with distance.

\subsection{Rosseland average interaction rate for the diffusive flux}

If the reduced MFP depends on energy, the energy-integrated true flux is given by Eq.~\eqref{eq:Ftrue-geometric} after performing the $\int d\omega$ integral. The diffusive flux, on the other hand, is given by the $\int d\omega$ integral of
Eq.~\eqref{eq:Fdiff-1}. In geometric variables, one finds
\begin{equation}
    F_{\rm diff}(z)=-\frac{1}{3}\int_{m_a}^\infty d\omega\,v_\omega^2\lambda_\omega(z)\frac{d}{dz}B_\omega(z)
    =-\frac{\nabla T}{3}\int_{m_a}^\infty d\omega\,\frac{v_\omega^3}{\Gamma_\omega}\frac{dB_\omega}{dT},
\end{equation}
where $B_\omega$, given in Eq.~\eqref{eq:B-source}, is the spectral blackbody density for a massless boson so that
\begin{equation}
    \frac{dB_\omega}{dT}=\frac{1}{2\pi^2}\,\frac{\omega^4 e^{\omega/T}}{T^2(e^{\omega/T}-1)^2}.
\end{equation}
For a massless boson with energy-independent MFP, the flux expression is
\begin{equation}\label{fig:Diffuse-Flux}
    F_{\rm diff}(z)=-\frac{\lambda}{3}\,\nabla T\int_{0}^\infty d\omega \frac{dB_\omega}{dT}
    =-\frac{\lambda}{3}\,\frac{2\pi^2}{15}\,T^3\nabla T
    =-\frac{\lambda}{3}\,\nabla B(z).
\end{equation}
Therefore, if we wish to express the general diffusive flux in terms of an equivalent average MFP, comparing the two expressions yields
\begin{equation}\label{eq:Rosseland}
    \lambda_{\rm eff}=\int_{m_a}^\infty d\omega\,\frac{v_\omega^3}{\Gamma_\omega}\,\frac{dB_\omega}{dT}
    \bigg/\int_{0}^\infty d\omega \frac{dB_\omega}{dT}
    =\frac{15}{4\pi^4}\,\frac{1}{T^5}\int_{m_a}^\infty d\omega\,
    \frac{\omega\,(\omega^2-m_a^2)^{3/2}\, e^{\omega/T}}{(e^{\omega/T}-1)^2\,\Gamma_\omega}.
\end{equation}
In radiative transport, this effective MFP corresponds to the Rosseland average of the interaction rate. 

\section{Boson luminosity in spherical geometry}

Our study is motivated by several recent papers concerning the FIB luminosity of a SN core and the associated energy loss. As we have argued in the previous section, the energy loss in the trapping limit can be estimated very well by quasi-thermal emission from a blackbody surface according to the Stefan-Boltzmann law. On the other hand, if one performs a numerical integration over an externally prescribed background model, one may as well use the exact expressions. Going beyond energy loss and asking for the nonlocal mode of energy transfer carried by FIBs, especially if these are radiatively unstable and can deposit energy far away from the point of production, a geometrically correct treatment is more important because a plane-parallel approximation is not 
appropriate if the energy is deposited far away from the compact stellar core. A similar question arises in the context of FIB energy loss and transfer in Horizontal Branch (HB) stars where the nonlocal transfer of energy was described as ``ballistic'' in contrast to that by diffusion \cite{Lucente:2022wai}. Therefore, we now turn to formulating the FIB flux in spherical geometry.

\subsection{Solution on a ray in geometric variables}

The solution for the stationary flux in any geometry derives from the stationary solution on a given ray of the radiation field that was discussed in Sec.~\ref{sec:StationaryState}. Because FIBs are only absorbed or emitted, but not scattered, different momentum modes of the FIB radiation field are decoupled and so a single ray provides the mother of all solutions. Following Sec.~\ref{sec:StationaryState} we thus consider a ray along some chosen FIB momentum direction, use a geometric coordinate~$s$, and express the solutions in terms of intensities instead of occupation numbers,
\begin{subequations}\label{eq:rayintensities}
\begin{eqnarray}
    I_\omega^+(s)&=&\frac{1}{v_\omega}\int_{-\infty}^s\! ds'\, Q_\omega(s')\,\exp\biggl[-\int_{s'}^{s}\frac{ds''}{\lambda_\omega(s'')}\biggr],
    \\
    I_\omega^-(s)&=&\frac{1}{v_\omega}\int_{s}^\infty ds'\, Q_\omega(s')\,\exp\biggl[-\int_{s}^{s'}\!\frac{ds''}{\lambda_\omega(s'')}\biggr],
\end{eqnarray}
\end{subequations}
where $\pm$ refers to the intensities of the FIB modes at the point $s$ along or opposite to the ray which has the direction of increasing $s$. The local FIB energy production rate $Q_\omega(s)=\Gamma_\omega(s) B_\omega(s)=v_\omega B_\omega(s)/\lambda_\omega(s)$ was defined earlier in Eq.~\eqref{eq:Qdefinition} and we assume that the blackbody intensity $B_\omega(s)$ and the reduced MFP $\lambda_\omega(s)$ are externally prescribed. The intensity at $s$ is simply the integral over the emission from downstream of the respective direction, modified by exponential damping along the way.

\subsection{Spherical volume integration: Observer perspective}

\begin{figure}[ht]
\centering
\includegraphics[width=0.35\textwidth]{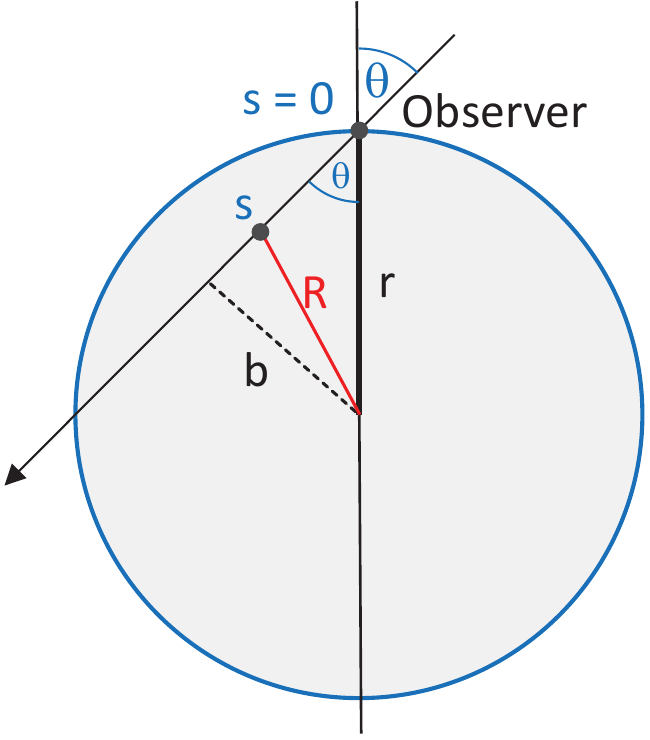}
  \caption{Geometric setting for calculating the boson flux at radius $r$ from the observer perspective.}\label{fig:VolumeIntegrals4}
\end{figure}

As a first of two ways to calculate the boson flux at radius $r$ in a spherically symmetric star we consider an observer at that radius and ask for the contribution of a given source to the outward or inward energy flux at that location and then integrate over all sources. Using the geometric setup shown in Fig.~\ref{fig:VolumeIntegrals4}, we consider a ray with a coordinate $s$ which is zero at $r$ and positive in the inward direction for later convenience. The ray is tilted relative to the radial direction by an angle $\theta$ with $\mu=\cos\theta$. Both $Q_\omega(r)$ and $\lambda_\omega(r)$ are assumed to be given as a function of stellar radius $r$. The impact parameter of this ray is $b=r\sin\theta$ and half the secant line is $a=r\cos\theta$. At point $s$ on this ray, the distance to the center of the star is given by $R^2=b^2+(a-s)^2$, providing
\begin{equation}\label{eq:rsmu}
  R_{r,s,\mu}=\sqrt{r^2+s^2-2rs\mu}.
\end{equation}
The intensities for the two directions on this ray at $s=0$ follow directly from Eq.~\eqref{eq:rayintensities}. 

However, we are interested in the energy flux in the radial direction, not the intensity, and so we need another factor $v_\omega\mu$, yielding at $s=0$ the energy fluxes
\begin{subequations}\label{eq:Fluxes-observer}
\begin{eqnarray}
    F_{\omega,\mu}^+(r)&=&\mu\int_{0}^\infty ds\, Q_\omega(R_{r,s,\mu})\,\exp\biggl[-\int_{0}^{s}\!\frac{ds'}{\lambda_\omega(R_{r,s',\mu})}\biggr],
    \label{eq:Fluxes-observer-a}
    \\
    F_{\omega,\mu}^-(r)&=&\mu\int_{-\infty}^0 \! ds\, Q_\omega(R_{r,s,\mu})\,\exp\biggl[-\int_{s'}^{0}\frac{ds'}{\lambda_\omega(R_{r,s',\mu})}\biggr].
\end{eqnarray}
\end{subequations}
Notice that by our choice of direction of the $s$ variable, it is $I^-_\omega$ that contributes to the radially outward flux $F^+_\omega$ and the other way around. Thus defined, both $F^\pm_{\omega,\mu}$ are positive if $\mu>0$ and the total flux is \smash{$F_{\omega,\mu}=F^+_{\omega,\mu}-F^-_{\omega,\mu}$}. To obtain the total flux, this expression must be integrated $\int_0^1 d\mu$ if we define the angle as the one between the ray and the radial direction as in Fig.~\ref{fig:VolumeIntegrals4}. However, as $F_{\omega,\mu}$ consists of a piece with positive and one with negative $\mu$, we may instead use only the first piece and integrate over all $\mu$ so that
\begin{equation}\label{eq:observer-flux}
    F_\omega(r)=
    \frac{1}2\int_{-1}^{+1}d\mu\,\mu\underbrace{\int_0^{\infty}\!ds\,Q_\omega(R_{r,s,\mu})\,
    \exp\left[-\int_0^s \frac{ds'}{\lambda_\omega(R_{r,s',\mu})}\right]}_{\displaystyle v_\omega I_{\omega,\mu}(r)}.
\end{equation}
The second integral is the local intensity $I_{\omega,\mu}(r)$ times the particle velocity $v_\omega$ as a function of direction.
The mono\-chromatic luminosity is $L_\omega(r)=4\pi r^2 F_\omega(r)$.

As a cross check we consider the strong trapping limit, where the particle MFP is short compared with $r$ and it makes sense to worry only about the region around $r$ with a few MFPs upstream and downstream. In this case, the general volume integral should approach the plane-parallel result. After integrating over emission angles, Eqs.~\eqref{eq:Fluxes-observer} are
\begin{subequations}\label{eq:Fluxes-observer-integrated}
\begin{eqnarray}
    F_{\omega}^+(r)&=&\frac{1}{2}\int_0^1 d\mu\,\mu\int_{0}^\infty ds\, Q_\omega(R_{s,\mu})\,\exp\biggl[-\int_{0}^{s}\!\frac{ds'}{\lambda_\omega(R_{s',\mu})}\biggr],
    \\
    F_{\omega}^-(r)&=&\frac{1}{2}\int_0^1 d\mu\,\mu\int_{-\infty}^0 \! ds\, Q_\omega(R_{s,\mu})\,\exp\biggl[-\int_{s}^{0}\frac{ds'}{\lambda_\omega(R_{s',\mu})}\biggr].
\end{eqnarray}
\end{subequations}
Because the $s$-integral contributes only for $s\ll R$ we may expand the expression Eq.~\eqref{eq:rsmu} for the radial position as
\begin{equation}
  R_{s,\mu}=r-s\mu+{\cal O}(s^2/r).
\end{equation}
The variable $s$ along the beam now always appears multiplied with $\mu$ and we introduce $z=-\mu s$, which is the radial distance to the position $r$. Notice that for positive $\mu$ a position at a radius larger than $r$ has negative $s$ by our convention for the direction of the considered ray. So the positive $z$-direction is the outward radiation direction and $R_{s,\mu}\to r+z$.

Because of strong trapping, regions a few MFPs upstream or downstream from $r$ are suppressed by the exponential damping factor, we may nominally extend the $dz$ integral to infinity. Therefore, the fluxes are
\begin{subequations}\label{eq:Fluxes-pp}
\begin{eqnarray}
    F_{\omega}^+(r)&=&\int_{-\infty}^0 \! dz\,Q_\omega(r+z)\,
    \frac{1}{2}\int_0^1 d\mu\,
    \exp\biggl[-\frac{1}{\mu}\int_{z}^{0}\frac{dz'}{\lambda_\omega(r+z')}\biggr],
    \\
    F_{\omega}^-(r)&=&\int_{0}^\infty dz\,Q_\omega(r+z)\,
    \frac{1}{2}\int_0^1 d\mu\,
    \exp\biggl[-\frac{1}{\mu}\int_{0}^{z}\frac{dz'}{\lambda_\omega(r+z')}\biggr].
\end{eqnarray}
\end{subequations}
The $d\mu$-integrals can now be expressed in terms of exponential integrals as explained around Eq.~\eqref{eq:expint}. The total flux $F_\omega=F_\omega^+-F_\omega^-$ can then be pieced together and reproduces a convolution integral in analogy to 
Eq.~\eqref{eq:Ftrue-geometric}. The analogy becomes perfect if we reinterpret $r$ as our $z$-variable and shift the integration variables accordingly.

\begin{figure}[b!]
\centering
\includegraphics[scale=0.50]{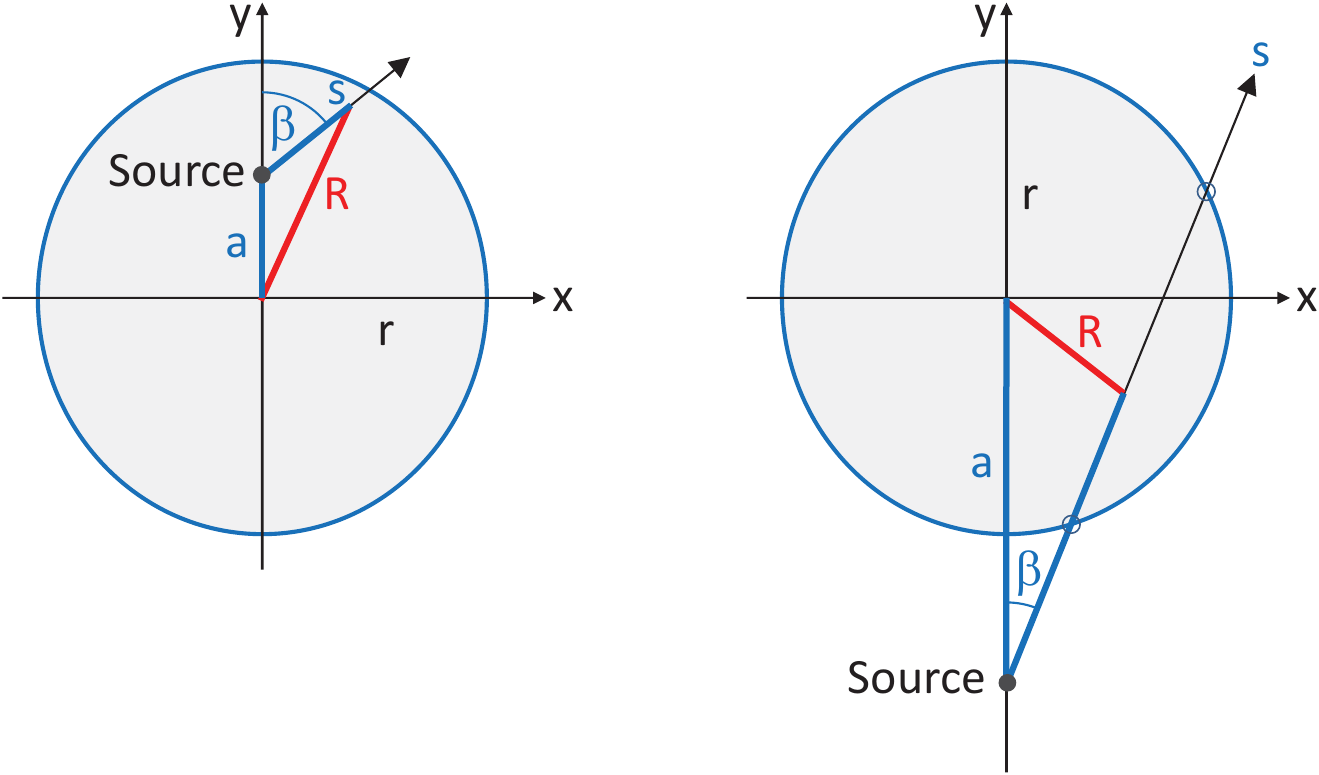}
  \caption{Geometric setting for calculating the contribution of a given source point to the energy flux at radius $r$.}\label{fig:VolumeIntegrals2}
\end{figure}

\subsection{Spherical volume integration: Source perspective}

We next turn to the second picture, sketched in Fig.~\ref{fig:VolumeIntegrals2}, where we consider emission from a given source and ask for its contribution to the outward and inward luminosities at some radius $r$. We begin with sources inside of $r$, all of which contribute to $L^+_\omega(r)$. We place the source at position $a<r$ on the $y$-axis, radiating isotropically in all directions characterized by the angle $\beta$. Following a ray in the direction $\beta$ with coordinate $s$ (origin at the source), the corresponding radius vector in the $x$-$y$-plane is ${\bf R}^<_{a,\beta,s}=(s\,\sin\beta,a+s\,\cos\beta)$ so that
\begin{equation}
  R^<_{a,\beta,s}=\sqrt{a^2+s^2+2sa\cos\beta}.
\end{equation}
By the same geometric consideration, the upper limit of integration is
\begin{equation}\label{eq:smax}
  s^{\rm max}_{r,a,\beta}=\sqrt{r^2-a^2\sin^2\beta}-a\cos\beta.
\end{equation}
Therefore, the intensity contribution of this source at radius $r$ is damped by
\begin{equation}
    \exp\biggl[-\int_{0}^{s^{\rm max}_{r,a,\beta}}\,\frac{ds}{\lambda_\omega({R^<_{a,\beta,s})}}\biggr].
\end{equation}
The ray punches through the $r$-sphere with an angle $\theta$ and so the outward flux requires a factor $\cos\theta$. On the other hand, if we think of the ray as having a small cross section, the area of intersection with the $r$-sphere is increased by $1/\cos\theta$ and so these two factors cancel. Actually if there were no damping of the emitted radiation, all bosons emitted from the source per unit time must pass the $r$-sphere and so the source contribution to the flux is the same, independently of the source position within the sphere. We only need to calculate the position-dependent average damping factor.
 Integrating over all source points within radius $r$,
we find
\begin{equation}\label{eq:Lplusinside}
    L^{+,{<}}_\omega(r)=\int_0^r\! da\,4\pi a^2\,Q_\omega(a)\,
    \frac{1}{2}\int_{-1}^{+1}d\cos\beta\,
    \exp\biggl[-\int_{0}^{s^{\rm max}_{r,a,\beta}}\,\frac{ds}{\lambda_\omega({R^<_{a,\beta,s})}}\biggr]
\end{equation}
for the outward luminosity provided by sources inside the sphere $r$. Notice that in this case, contrary to the observer perspective, it is not possible to uniquely define the flux and we therefore work with the luminosity. The two quantities are however easy related $ F^{+,{<}}_\omega(r)= L^{+,{<}}_\omega(r)/4\pi r^2$.

For a source outside the $r$-sphere (right panel in Fig.~\ref{fig:VolumeIntegrals2}), a ray passes through the sphere if the angle $\beta$ is constrained by $\sin\beta<r/a$ so that
\begin{equation}
    c^{\rm min}_{r,a}=\cos\beta_{\rm max}=\sqrt{1-r^2/a^2}.
\end{equation}
The distance from the stellar center of a point $s$ on the ray is
\begin{equation}
  R^>_{a,\beta,s}=\sqrt{a^2+s^2-2sa\cos\beta}.
\end{equation}
The length on the ray until the first and second points of intersection is
\begin{equation}
    s^{(1,2)}_{r,a,\beta}=a\cos\beta\pm \sqrt{r^2-a^2\sin^2\beta}.
\end{equation}
For the flux contribution of a ray intersecting a tilted surface, the same remarks apply as earlier. The first intersection point contributes to the inward flux, the second one to the outward flux. Collecting everything, we find for the total flux
\begin{equation}\label{eq:source-flux}
    L_\omega(r)=\int_0^\infty\! da\,4\pi a^2\,Q_\omega(a)\,{\sf E}_\omega(r,a),
\end{equation}
where for $a<r$
\begin{equation}
    {\sf E}^<_\omega(r,a)=\frac{1}{2}\int_{-1}^{+1}d\cos\beta\,
    \exp\biggl[-\int_{0}^{s^{\rm max}_{r,a,\beta}}\,\frac{ds}{\lambda_\omega({R^<_{a,\beta,s})}}\biggr]
\end{equation}
and for $a>r$
\begin{equation}
    {\sf E}^>_\omega(r,a)=\frac{1}{2}\int_{c^{\rm min}_{r,a}}^{+1}d\cos\beta\,\left\lbrace
    \exp\biggl[-\int_{0}^{s^{(2)}_{r,b,\beta}}\,\frac{ds}{\lambda_\omega({R^>_{a,\beta,s})}}\biggr]
    -\exp\biggl[-\int_{0}^{s^{(1)}_{r,b,\beta}}\,\frac{ds}{\lambda_\omega({R^>_{a,\beta,s})}}\biggr]
    \right\rbrace
\end{equation}
and ${\sf E}_\omega(r,a)=0$ for $r=a$.

While this result looks far more complicated than the one found from the observer perspective, its structure is more reminiscent of the plane-parallel case in that we convolve the radial source distribution $Q_\omega(r)$ with the kernel ${\sf E}_\omega(r,a)$ which is positive for $a<r$ and negative for $a>r$ and the local flux is determined by the sources a few MFPs inside and outside the considered radius.

The formal transition to the plane-parallel case is made by assuming the MFP is so small that the $a$-integration contributes only in a thin shell around $r$ and we set $a=r+z$ with $|z|\ll r$. The angle integration for $z<0$ covers only the range $0<\cos\beta<1$ and so for $z<0$ we find
\begin{equation}
    {\sf E}^<_\omega(r,z)=\frac{1}{2}\int_{0}^{1}d\cos\beta\,
    \exp\biggl[-\int_{0}^{-z/\cos\beta}\,\frac{ds}{\lambda_\omega(r+z-s\cos\beta)}\biggr].
\end{equation}
Substituting $z'=-s\cos\beta$, this becomes
\begin{equation}
    {\sf E}^<_\omega(r,z)=\frac{1}{2}\int_{0}^{1}d\cos\beta\,
    \exp\biggl[-\frac{1}{\cos\beta}\int_{z}^{0}\,\frac{dz'}{\lambda_\omega(r+z')}\biggr]
    =\frac{1}{2}E_2\biggl[\int_{z}^{0}\,\frac{dz'}{\lambda_\omega(r+z')}\biggr].
\end{equation}
The derivation for $z>0$ is analogous if we notice that the contribution from the second intersection point can be dropped in the present limit. Therefore, ${\sf E}_\omega(r,a)$ in the small-$\lambda$ limit is pieced together from exponential integral functions as in the plane-parallel case.

\subsection{Luminosity at infinity}

Another important limit is the FIB luminosity at infinity, corresponding to the total energy loss of the star or SN core in the form of FIBs. This quantity is only useful when the FIBs are essentially stable, otherwise the flux at infinity always vanishes. In practice, we take $r$ to be much larger than the geometric size of the production region, but much smaller than the MFP against decay. As a consequence, there are no sources outside the very large $R$, so the large-$R$ limit can be taken on the basis of the outward luminosity.

Beginning with the ``observer perspective,'' the starting point is the outgoing flux of Eq.~\eqref{eq:Fluxes-observer-a}, implying a luminosity
\begin{equation}
  L_{\omega,\mu}^+(r)=4\pi r^2\,\frac{1}{2}\int_0^1d\mu\,\mu\int_{0}^\infty ds\, Q_\omega(R_{r,s,\mu})\,\exp\biggl[-\int_{0}^{s}\!\frac{ds'}{\lambda_\omega(R_{r,s',\mu})}\biggr].
\end{equation}
By assumption, regions far away from the star do not contribute to FIB production or decay, so the range of angles $\theta$ that contribute become infinitesimally small for $r\to\infty$. This singularity is avoided by using instead the impact parameter $b=r\sin\theta$ as integration variable. Moreover, the integration along the ray shown in Fig.~\ref{fig:VolumeIntegrals4} is performed in a shifted variable $z=s-r$, which amounts in the limit $r\to\infty$ to putting the zero-point of this variable at the point of intersection of the impact line $b$ with the ray. Using these coordinates, the radial position is $R_{r,s,\mu}\to R_{z,b}=(b^2+z^2)^{1/2}$. The angle integral thus becomes
\smash{$4\pi r^2 \frac{1}{2}\int_0^1 d\mu\,\mu \ldots\to 2\pi \int_0^\infty db\, b \ldots$},
where we have used $\mu=1$, $b=r\sin\theta=r\theta$, and we have extended the $b$-integration to $\infty$ because only regions with $b\ll r$ contribute by assumption. Notice that $2\pi b$ is the circumference of a circle with radius $b$, so the $b$-integration is simply one over the stellar disk in terms of the radius (or impact parameter) on the disk. The volume integration has become one over the stellar disk and, transverse to it in the observer direction, over the new variable $z$. With  $L_\omega=L_\omega^+(\infty)$, collecting everything, and re-naming the variable of integration $b\to r$ we find
\begin{equation}\label{eq:Linfty-observer}
    L_\omega=\int_0^\infty \!dr\,2\pi r\int_{-\infty}^{+\infty} dz\,
    Q_\omega\bigl(\sqrt{r^2+z^2}\bigr)
    \,\exp\biggl[-\int_z^{\infty}\!\frac{dz'}{\lambda_\omega\bigl(\sqrt{r^2+z^{\prime\,2}}\bigr)}\biggr].
\end{equation}
The total energy loss finally requires an integration over $d\omega$.

We next turn to the ``source perspective'' and note that for $r\to\infty$ all sources are within the $r$-sphere, so as a starting point we may use Eq.~\eqref{eq:Lplusinside} for the outward flux caused by sources within the $r$-sphere. In the limit $r\to\infty$ the upper limit of integration becomes $s^{\rm max}_{R,a,\beta}\to\infty$. Collecting everything and re-naming the variable of integration $a\to r$
\begin{equation}\label{eq:Linfty-source}
    L_\omega=\int_0^\infty\! dr\,4\pi r^2\,Q_\omega(r)\,
    \underbrace{\frac{1}{2}\int_{-1}^{+1}\!d\cos\beta\,
    \exp\biggl[-\int_{0}^{\infty}\,\frac{ds}{\lambda_\omega\bigl(\sqrt{r^2+s^2+2rs\cos\beta}\bigr)}\biggr]}_{
    \displaystyle {\sf T}_\omega(r)=\bigl\langle e^{-\tau_{\omega,\mu}(r)}\bigr\rangle_{\rm angles}}.
\end{equation}
The intuitive meaning is that we perform a volume integral over the radius-dependent energy-loss rate, reduced by the angle-averaged transmittance ${\sf T}_\omega(r)$, where $\tau_{\omega,\mu}(r)$ is the optical depth of the source point in a specific direction of emission.

The expressions for $L_\omega$ from the observer perspective Eq.~\eqref{eq:Linfty-observer} and from the source perspective Eq.~\eqref{eq:Linfty-source} are both intuitive, yet look very different. However, one can show with a direct transformation of the integral expressions that they are indeed the same.

\subsection{Transmittance in the strong-trapping limit}

To calculate the FIB flux at infinity in spherical symmetry, the crucial geometric information in Eq.~\eqref{eq:Linfty-source} is encoded in the angle-averaged transmittance 
\begin{equation}\label{eq:transmittance}
    {\sf T}_\omega(r)=
    \frac{1}{2}\int_{-1}^{+1}\!d\mu\,
    \exp\biggl[-\int_{0}^{\infty}\,\frac{ds}{\lambda_\omega\bigl(\sqrt{r^2+s^2+2rs\mu}\bigr)}\biggr],
\end{equation}
where we have renamed $\cos\beta\to\mu$. The integral in the exponential is the optical depth $\tau_{r,\mu}$ for a FIB emitted at radius $r$ with direction $\mu=\cos\beta$ relative to the radial direction. 

In some recent papers \cite{Bollig:2020xdr, Croon:2020lrf}, the transmittance was estimated as $e^{-\tau(r)}$, where $\tau(r)$ is
the optical depth in the outward-radial direction ($\beta=0$), i.e., the shortest way out. This approximation overestimates the transmittance except at the center of the star because otherwise the optical depth is larger in all directions compared with the radial-outward one. The prescription of Ref.~\cite{Chang:2016ntp} and re-used in Ref.~\cite{Lucente:2020whw} effectively employs an even larger transmittance. For a specific SN model, the difference between the naive transmittance  $e^{-\tau(r)}$ and that of Eq.~\eqref{eq:transmittance} is shown in Fig.~\ref{fig:L-Primakoff} below (dashed vs.\ solid blue line).

While there is no general answer concerning the difference, it is easy to estimate in the strong-trapping limit where the FIBs essentially emerge from a Stefan-Boltzmann sphere at $\tau(r_{\rm SB})=2/3$ and if we assume that the absorption rate decreases fast with radius at and beyond $r_{\rm SB}$. So if the geometric atmospheric height of the decoupling region is small relative to the decoupling radius we are back to the plane-parallel atmosphere approximation. In Eq.~\eqref{eq:transmittance} this implies that $s\ll r$ for the contributing range, implying that $\sqrt{r^2+s^2+2rs\mu}\to r+s\mu$. As a variable of integration we choose the vertical depth $z=\mu s$ and we also note that in the strong-trapping limit the transmittance for inward-bound directions vanishes, so the $d\mu$ integral is only over positive $\mu$. Collecting everything, we find in the plane-parallel approximation
\begin{equation}\label{eq:transmittance-2}
    {\sf T}(\tau)=
    \frac{1}{2}\int_{0}^{1}\!d\mu\,e^{-\tau/\mu}=\frac{1}{2}\,E_2(\tau),
\end{equation}
where $E_2(\tau)$ is the second exponential integral defined in Eq.~\eqref{eq:expint}. In the plane-parallel case, the transmittance only depends on optical depth, not geometric radial position, where here $\tau$ stands for the ``outward radial'' optical depth of a given source. We have also dropped the index $\omega$ for convenience.

We may compare ${\sf T}(\tau)$ with the naive value $e^{-\tau}$ in various cases. For $\tau=0$ we have $e^{-\tau}=1$ and $E_2(\tau)=1$ and so ${\sf T}(0)=1/2$, which is 1/2 times the naive value of~1. The reason is that of all bosons launched at the surface, only the outward-moving ones escape. For very large $\tau$, $E_2(\tau)/\exp(-\tau)\to \tau^{-1}$, so besides the previous factor 1/2 concerning the inward-bound bosons, the naive transmittance is $\tau$ times the true one and thus a vast overestimate because any trajectory that deviates only mildly from the exact radial direction implies much larger absorption. Finally, for the Stefan-Boltzmann value $\tau=2/3$, the ratio is 0.4968. In absolute terms, ${\sf T}(2/3)=0.1239$, meaning that around 1 in 8 FIBs produced at $\tau=2/3$ makes it to infinity. Counting only the outward-bound ones ($\mu>0$), almost exactly one in four escapes.

\section{Explicit example I: Supernova energy loss through Primakoff production}

As a first example we consider axion-like particles (ALPs) with a generic two-photon interaction encoded in the coupling strength $G_{a\gamma\gamma}$ and with a mass so small that decays are irrelevant and that we can use ultrarelativistic kinematics. In this case they are absorbed or produced only by the Primakoff process $\gamma+Ze\leftrightarrow Ze+a$ on charged particles. It is reasonable to approximate the reduced absorption rate $\Gamma_\omega$ as independent of energy \cite{Caputo:2021rux}, so this case comes close to the ``grey atmosphere'' approximation of radiative transfer theory. We use our expressions to calculate the total energy-loss rate $L_a$ for a prescribed numerical SN model as a function of $G_{a\gamma\gamma}$, compare it with the neutrino luminosity $L_\nu$, and find the two solutions for $G_{a\gamma\gamma}$ where $L_a=L_\nu$. The trapping solution is found to agree very well with the one from the Stefan-Boltzmann argument as anticipated.

\subsection{Interaction model}
\label{sec:PrimakoffInteractionModel}

We now consider massless ALPs that are assumed to interact with the electromagnetic field through the Lagrangian
\begin{equation}
    {\cal L}_{a\gamma\gamma}=- \, G_{a\gamma\gamma}\frac{a}{4}\, F_{\mu\nu}\tilde F^{\mu\nu}=G_{a\gamma\gamma}\,a\,{\bf E}\cdot{\bf B},
\end{equation}
where $G_{a\gamma\gamma}$ is a coupling constant with dimension (energy)$^{-1}$. It is the only particle-physics parameter entering our discussion. ALPs are dominantly absorbed by the Primakoff process $a+Ze\to Ze+\gamma$ on charged particles with a rate
\begin{equation}
    \Gamma_{\rm A}=Z^2\alpha G^2_{a\gamma\gamma}n_Z f_{\rm S} f_{\rm B},
\end{equation}
where $n_Z$ is the number density of targets, $f_{\rm S}$ a screening factor, and $f_{\rm B}$ a Bose stimulation factor for the final-state photon. The rate has been summed over final-state photon polarizations.

An exact evaluation of this rate for the conditions of a SN core is not available because there are many complications as detailed in Sec.~II.E of Ref.~\cite{Caputo:2021rux}. Electrons as targets are relativistic and degenerate and will be neglected. Charged nuclear targets are not only protons (as had often been assumed), but in the hottest and most important regions also small nuclear clusters. Neglecting electrons one can use  Debye-H\"uckel screening \cite{Raffelt:1985nk}, but here as well as in the target phase space we neglect degeneracy effects, probably not a bad approximation in the relevant hottest regions. As suggested in Ref.~\cite{Caputo:2021rux} we finally set $\sum Z^2 n_Z\to (1-Y_n)n_{\rm B}$, where $Y_n$ is the neutron abundance (number of neutrons per baryon), keeping in mind that in general $1-Y_n$ is {\em not} the same as the proton abundance, although we call it the effective proton abundance. The screening factor varies only slowly in the range of energy relative to the screening scale and, given the relatively rough approximations used, we may as well set it to unity. Finally, the Bose stimulation factor is $f_{\rm B}=(1+f_\gamma)$ and because the targets do not recoil much, the photon energy is nearly the same as the ALP energy $\omega$, so $f_\gamma=1/(e^{\omega/T}-1)$ and we note that $f_{\rm B} = 1+f_\gamma=1/(1-e^{-\omega/T})$. Multiplication with $(1-e^{-\omega/T})$ to obtain the reduced absorption rate and collecting everything yields for the latter
\begin{equation}
  \Gamma=\underbrace{\alpha G_{a\gamma\gamma}^2}_{\displaystyle \sigma_a}~
  \underbrace{(1-Y_n)\,n_{\rm B}}_{\displaystyle \hat n}.
\end{equation}
Numerically, the cross section is
\begin{equation}\label{eq:cross-section}
  \sigma_a=2.84 \times 10^{-42}~{\rm cm}^2~G_6^2,
  \quad\hbox{where}\quad
  G_6=\frac{G_{a\gamma\gamma}}{10^{-6}\,{\rm GeV}^{-1}}.
\end{equation}
In this way we are naturally led to the simple case of a grey-atmosphere model which is defined by the reduced absorption rate not to depend on energy. In this case the energy integral in Eq.~\eqref{eq:Linfty-source} can be done explicitly and the ALP luminosity at infinity is
\begin{equation}\label{eq:spherical-integral-Primakoff}
  L_a=\int_{0}^{\infty}\!\!dr\,\underbrace{4\pi r^2\,B(r)\,\sigma_a \hat n(r)}_{\displaystyle L'_a(r)}
  \,{\sf T}(r),
\end{equation}
where the angle-averaged transmittance ${\sf T}(r)$ following from Eq.~\eqref{eq:transmittance} is
\begin{equation}\label{eq:transmittance-Primakoff}
    {\sf T}(r)=
    \frac{1}{2}\int_{-1}^{+1}\!d\mu\,
    \exp\biggl[-\int_{0}^{\infty}\,ds\,\sigma_a\hat n\bigl(\sqrt{r^2+s^2+2rs\mu}\bigr)\biggr].
\end{equation}
All we need to evaluate $L_a$ is a profile of $(1-Y_n)\rho_{\rm B}$ and of the temperature.

\subsection{Supernova model and its ALP flux}

\begin{figure}[ht]
\hbox to\textwidth{\hfill\includegraphics[width=0.4\textwidth]{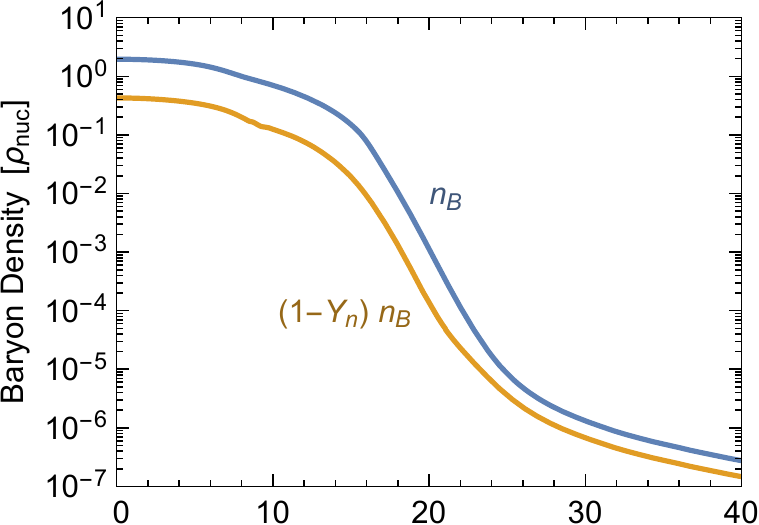}\kern20pt\includegraphics[width=0.4\textwidth]{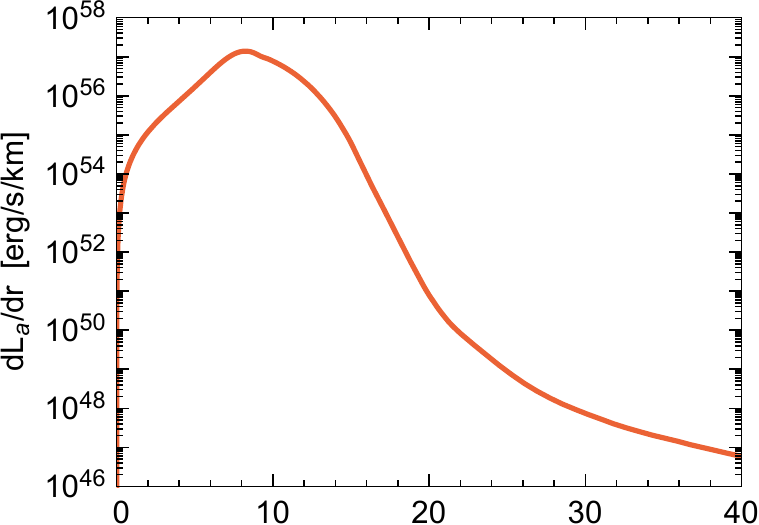}\hfill}
\hbox to\textwidth{\hfill\includegraphics[width=0.4\textwidth]{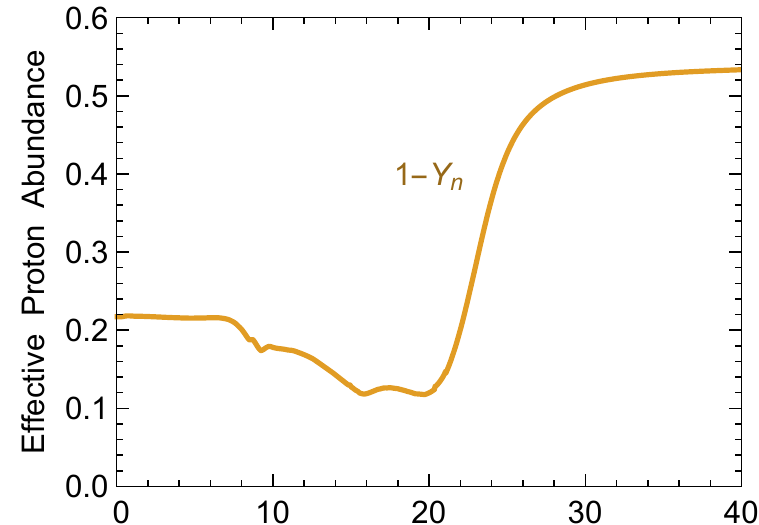}\kern20pt\includegraphics[width=0.4\textwidth]{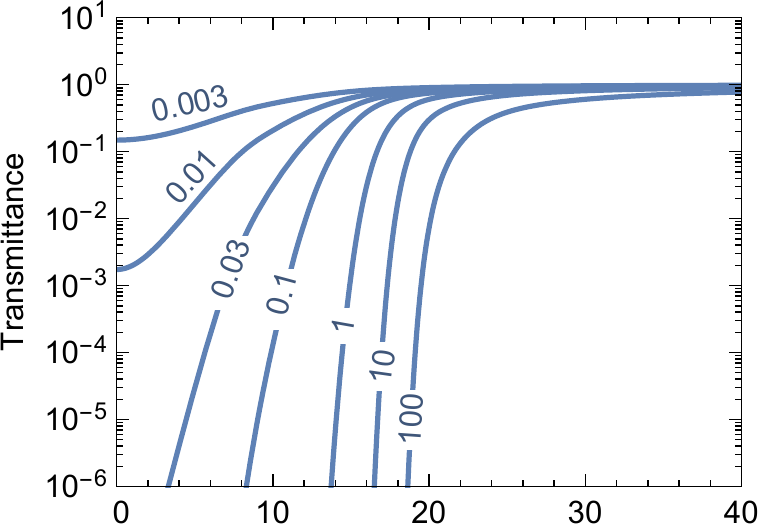}\hfill}
\hbox to\textwidth{\hfill\includegraphics[width=0.4\textwidth]{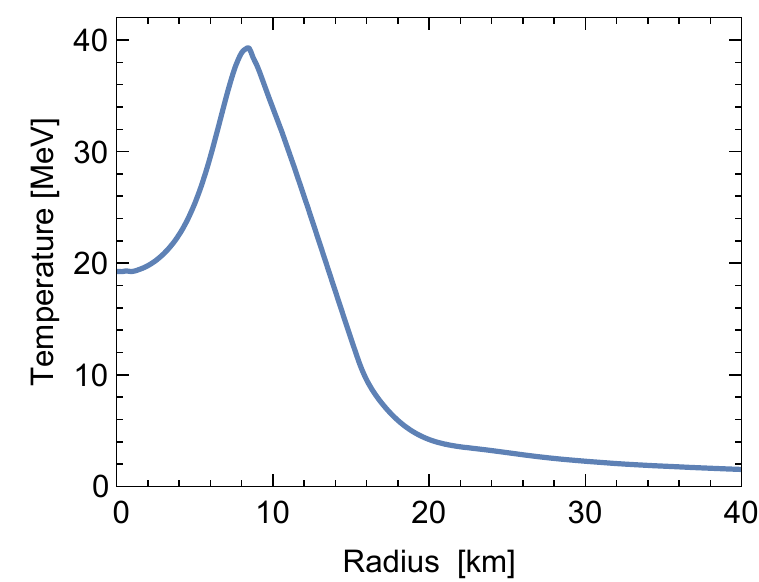}\kern20pt\includegraphics[width=0.4\textwidth]{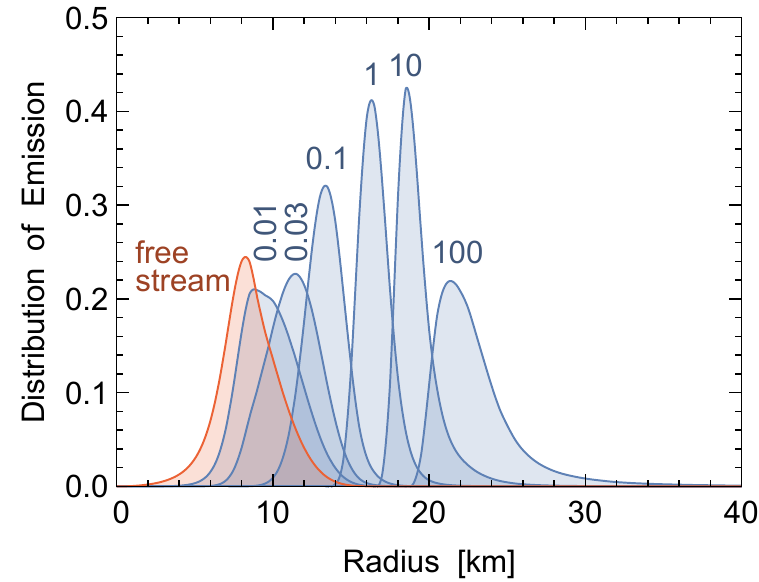}\hfill}
  \caption{Supernova model described in the text. {\em Left column:} Baryon density (in terms of nuclear density), effective proton abundance, and temperature as indicated.   {\em Right column:} ALP production distribution $L'_a(r)$ in the top panel is for $\sigma_a=10^{-41}\,{\rm cm}^2$ ($\sigma_{41}=1$). On a linear scale it corresponds to the red curve in the bottom panel. The transmittance is shown for the indicated values of $\sigma_{41}$. The bottom panel shows the normalized distributions $L'_a(r)$ for the indicated values of~$\sigma_{41}$.}\label{fig:SN-Model}
\end{figure}

To illustrate these results we evaluate them explicitly for a numerical SN model. We use the Garching muonic SN model SFHo-18.8 at $t_{\rm pb}=1$\,s that was used in several recent studies of SN particle bounds \cite{Bollig:2020xdr,Caputo:2021rux,Caputo:2022mah}. These SN models include muons, which is a generic physical effect, although not crucial for our discussion. The models are spherically symmetric, but include convection in the form of a mixing-length treatment. (The fixed $T$ gradient in the approximate range 8--15~km seen in the left-middle panel of Fig.~\ref{fig:SN-Model} reflects convection.) The final neutron-star baryonic mass is $1.351\,M_\odot$, the final gravitational mass is $1.241\,M_\odot$, so the total amount of liberated gravitational binding energy is the difference which is $1.98\times 10^{53}\,{\rm erg}$. Therefore, the released neutrino energy is at the lower end of the typical range, whereas the duration of neutrino emission is relatively short (due to convection) and the maximum temperature of around 40~MeV reached in the core is relatively small. More details are shown in Refs.~\cite{Bollig:2020xdr,Caputo:2021rux}, whereas the parameters relevant for us are plotted in Fig.~\ref{fig:SN-Model}.

To calculate the ALP luminosity, in principle one should include gravitational effects that are also included in numerical SN models, notably gravitational redshift as outlined in Refs.~\cite{Caputo:2021rux,Caputo:2022mah}. On the other hand, our entire treatment of radiative transfer has ignored such effects and in particular redshift and bending of trajectories. Here we are not performing a precision analysis of particle bounds but rather illustrate the relationship between volume-emission and boson-sphere Stefan-Boltzmann emission. Therefore we continue to ignore gravitational effects.

We express the ALP interaction strength in terms of the cross section Eq.~\eqref{eq:cross-section} that we parameterized in terms of $\sigma_{41}=\sigma_a/10^{-41}\,{\rm cm}^2$. The scale is chosen such that for $\sigma_{41}\simeq 1$ the ALP sphere will be close to the neutrino sphere at a radius of around 17~km. In the right-top panel of Fig.~\ref{fig:SN-Model} we show the ALP production rate $L_a'(r)$ for $\sigma_{41}=1$ defined in Eq.~\eqref{eq:spherical-integral-Primakoff}. In  normalized form and on a linear vertical scale it is the same as the red curve in the right-bottom panel. The maximum of emission is near the $T$ maximum. In addition, the central stellar region is geometrically suppressed by the $4\pi r^2$ factor.

In the right-middle panel, we show the transmittance of Eq.~\eqref{eq:transmittance-Primakoff} for the indicated values of $\sigma_{41}$, whereas in the right-bottom panel we show the product $L_a'(r){\sf T}(r)$ in normalized form, i.e., the source distribution of the escaping ALPs. We see that the ALPs always originate from a shell of thickness of a few km. In the free-streaming limit (unit transmittance) this shell is simply given by the product of the $T^4$ and $\hat n$ profiles together with the geometric $4\pi r^2$ factor. For larger coupling strengths, the emitting shell moves outward, driven by the transmittance that steeply falls for smaller radius, and the production rate, that steeply falls for larger $r$. However, the resulting shell is never very thin. The variation of widths of these normalized curves is also represented by their variation in height and we glean from the plot that the radial region of emission becomes less than a factor of 2 sharper for ``surface emission'' instead of free-streaming volume emission. This conclusion agrees with the schematic plane-parallel atmosphere model shown in Fig.~\ref{fig:SourceDistribution}.

\begin{figure}[ht]
\centering
\includegraphics[width=0.4\columnwidth]{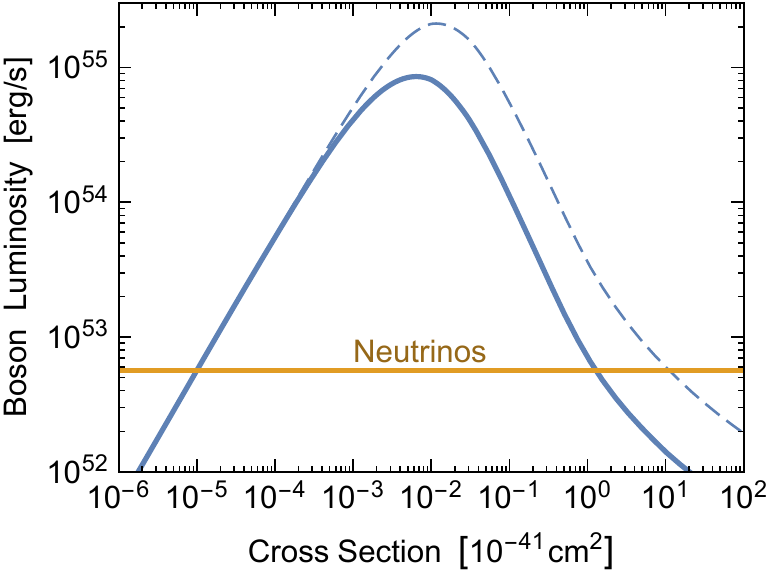}
\caption{ALP luminosity for our unperturbed SN model as a function of the effective Primakoff cross section on protons, to be compared with the neutrino luminosity. The blue solid line uses the full transmittance of Eq.~\eqref{eq:transmittance-Primakoff}, whereas the dashed line uses the naive transmittance $e^{-\tau(r)}$; here $\tau(r)$ is optical depth in the outward radial direction.}\label{fig:L-Primakoff}
\end{figure}

Next we show in Fig.~\ref{fig:L-Primakoff} the ALP luminosity thus derived as a function of the Primakoff cross section. We compare it with the neutrino luminosity $L_\nu=5.68\times10^{52}\,{\rm erg/s}$ of this model. This value corresponds approximately to the neutrino-sphere region around 17\,km, whereas after taking redshift effects into account it is $L_\nu=4.4\times10^{52}\,{\rm erg/s}$ for an observer at infinity. However, as we do not include redshift effects in our ALP luminosity calculation, we compare the luminosities roughly in the local environment.

On the free-streaming side, the two luminosities are equal for $\sigma_a=1.0\times10^{-46}\,{\rm cm}^2$, corresponding to $G_{a\gamma\gamma}=0.59\times10^{-10}\,{\rm GeV}^{-1}$. On the trapping side, they are equal for $\sigma_a=1.27\times10^{-41}\,{\rm cm}^2$, corresponding to $G_{a\gamma\gamma}=2.1\times10^{-6}\,{\rm GeV}^{-1}$. In which sense these $G_{a\gamma\gamma}$ values should be seen as constraints has been discussed elsewhere \cite{Caputo:2021rux}. Here we simply take them as the values where the ALP luminosity, calculated on an unperturbed SN model, equals $L_\nu$ of that model.

In the trapping limit, we may compare the ALP flux with the one found from the Stefan-Boltzmann argument. In our model, the SB flux $4\pi r_{\rm SB}^2 (\pi^2/120)T_{\rm SB}^4$ equals $L_\nu$ for $r_{\rm SB}=16.99\,{\rm km}$. The cross section required to achieve $\tau=2/3$ at this radius is $\sigma_a=1.03\times10^{-41}\,{\rm cm}^2$, corresponding to
$G_{a\gamma\gamma}=1.9\times10^{-6}\,{\rm GeV}^{-1}$. Therefore, within 10\% one finds the same coupling strength as one found with the full transmittance-modified volume integration. The errors incurred by all other approximations, for example concerning the Primakoff cross section and concerning the impact of gravity, are of similar magnitude. Therefore, on this level of precision there is no particular benefit in performing the full volume integration that can be numerically cumbersome. 

Notice that using the transmittance $e^{-\tau(r)}$ with the optical depth only in the outward-radial direction (dashed line in Fig.~\ref{fig:L-Primakoff}) would lead, for the trapping regime, to the bound $G_{a\gamma\gamma}=6.1\times10^{-6}\,{\rm GeV}^{-1}$, a factor 3 more stringent than the correct one. This further stresses the importance of considering the correct angle-averaged transmittance as already discussed around Eq.~\eqref{eq:transmittance-2}.

\subsection{Energy transfer by ALPs}

Besides the SN energy loss (the luminosity seen by a distant observer) we may also ask for the ALP flux $L_a(r)$ as a function of radius in and near the SN. Its radial variation reveals the energy gain or loss by the local medium caused by ALP emission and absorption. In the source-perspective expression of Eq.~\eqref{eq:source-flux} follows that the kernel ${\sf E}_\omega(r,a)$, in our present case, does not depend on $\omega$ and only on the radial variation of the MFP that here does not depend on temperature, so the kernel depends only on $\hat n(r)$ and the chosen value of $\sigma_a$.

\begin{figure}[ht]
\centering
\includegraphics[width=0.4\textwidth]{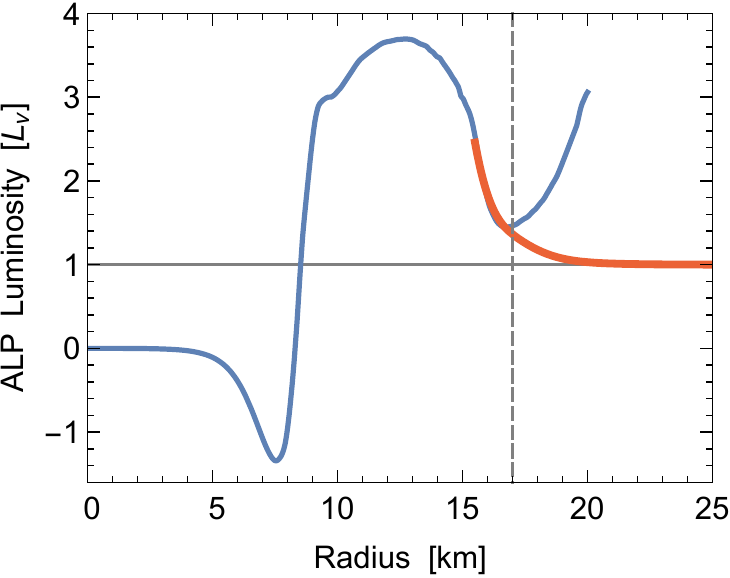}
\caption{Radial variation of ALP luminosity in our SN model for $\sigma_{41}=1.27$. {\em Blue line}: Flux predicted in the diffusion approximation. {\em Red line}: True flux based on Eq.~\eqref{eq:source-flux}. The Stefan-Boltzmann radius of 17.0\,km, where $\tau=2/3$, is marked with a vertical dashed line.}\label{ALPflux-radial}
\end{figure}

For illustration we use the trapping limit and specifically $\sigma_{41}=1.27$, where the ALP flux at infinity matches $L_\nu=5.68\times10^{52}\,{\rm erg/s}$. In Fig.~\ref{ALPflux-radial} we show as a blue line the radial flux variation based on the diffusion approximation. As an orange line we show the true flux based on
Eq.~\eqref{eq:source-flux}. The two curves separate in the decoupling region around 17\,km where $\tau=2/3$. Beyond this region, the true ALP flux is constant. Deeper inside, it agrees with the diffusive result. We see that for radii smaller than the decoupling region, ALPs carry a significant energy flux and so would play a significant role for energy transfer within the star.

\section{Explicit example II: Two-photon decay and photon coalescence}

As a second explicit case we consider ALPs with a mass $m_a$ so large that photon coalescence $2\gamma\to a$ is the main production process, not Primakoff production which we now ignore. In a SN core, this situation pertains for $m_a\gtrsim 60\,{\rm MeV}$ \cite{Lucente:2020whw} or in the core of horizontal-branch stars for $m_a\gtrsim 50\,{\rm keV}$ \cite{Lucente:2022wai}. In this situation, the only information from the stellar model is the temperature profile, whereas for the ALP both the coupling strength and the mass enter.

\subsection{Interaction model}

Once more we consider generic ALPs with a two-photon coupling discussed in Sec.~\ref{sec:PrimakoffInteractionModel}. In the ALP rest frame, the two-photon decay rate is
\begin{equation}
    \Gamma_a=\frac{G_{a\gamma\gamma}^2m_a^3}{64\pi},
\end{equation}
which we use as our primary parameter to quantify the interaction strength. 

The ``absorption'' rate caused by the decay $a\to2\gamma$ for pseudoscalar FIBs was explicitly provided in the Supplementary Material of Ref.~\cite{Caputo:2022mah}. Starting from their Eqs.~(S10) and (S11), the reduced absorption rate is
\begin{equation}\label{eq:Gamma-decay}
    \Gamma_\omega=\Gamma_a\,\frac{m_a}{\omega}\,g_{\rm B}(\omega),
    \quad\hbox{where}\quad
    g_{\rm B}(\omega)=\frac{2T}{v_\omega\omega}
     \log\frac{\sinh\frac{(1+v_\omega)\,\omega}{4T}}{\sinh\frac{(1-v_\omega)\,\omega}{4T}}
\end{equation}
and $v_\omega=(1-m_a^2/\omega^2)^{1/2}$ is the boson velocity. Here $g_{\rm B}$ accounts for final-state Bose stimulation in the decay. Compared with the factor $f_{\rm B}$ of Ref.~\cite{Caputo:2022mah}, $g_{\rm B}$ includes $(1-e^{-\omega/T})$ for the {\em reduced} absorption rate. In the limit $T\to0$ it is $g_{\rm B}\to1$ and we are back to the vacuum decay rate. The boson flux arising in Eq.~\eqref{eq:Ftrue-geometric} is here physically produced by photon coalescence $2\gamma\to a$, a process encoded in the reduced absorption rate of Eq.~\eqref{eq:Gamma-decay}. In particular, the temperature of the background medium enters only through $g_{\rm B}$.

The local energy production rate in the form of ALPs is $B_\omega v_\omega \Gamma_\omega$ or explicitly
\begin{equation}
    Q_\omega=\frac{\Gamma_a}{\pi^2}\,\frac{m_a\omega T}{e^{\omega/T}-1}\,
    \log\frac{\sinh\frac{\omega+\sqrt{\omega^2-m_a^2}}{4T}}{\sinh\frac{\omega-\sqrt{\omega^2-m_a^2}}{4T}},
\end{equation}
for example in units of ${\rm erg}\,{\rm cm}^{-3}\,{\rm s}^{-1}\,{\rm MeV}^{-1}$.

\subsection{Diffusive energy transfer}

\begin{figure}[ht]
\centering
\includegraphics[width=0.4\textwidth]{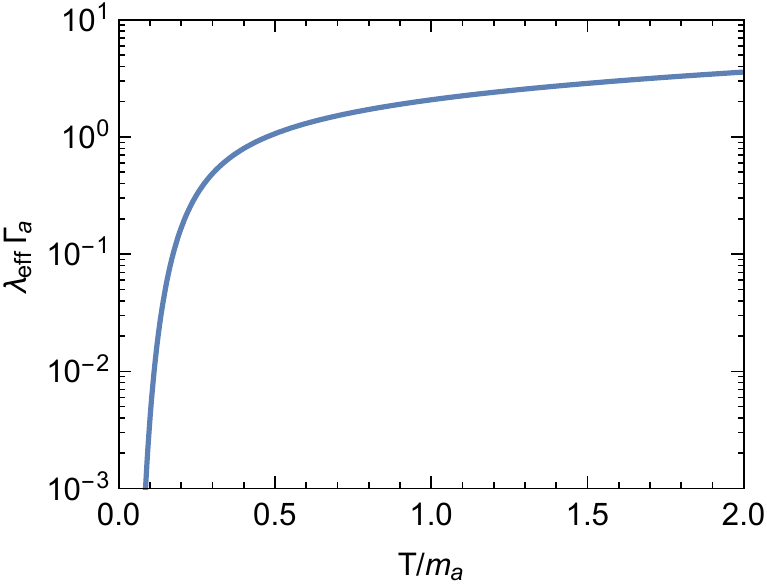}
\vskip-6pt
\caption{Effective mean-free path according to Eq.~\eqref{eq:lambda-eff}.}\label{fig:Lambda}
\end{figure}

To calculate the luminosity $L_\omega(r)$ in Eq.~\eqref{eq:source-flux} we need the MFP, which in our case is explicitly
\begin{equation}
    \frac{1}{\lambda_\omega}=\frac{\Gamma_\omega}{v_\omega}
    =\Gamma_a\,\frac{2 m_a T}{\omega^2-m_a^2}\,\log\frac{\sinh\frac{\omega+\sqrt{\omega^2-m_a^2}}{4T}}{\sinh\frac{\omega-\sqrt{\omega^2-m_a^2}}{4T}}.
\end{equation}
According to Eq.~\eqref{eq:Rosseland}, the Rosseland average for the effective MFP is
\begin{equation}\label{eq:lambda-eff}
    \lambda_{\rm eff}=\frac{1}{\Gamma_a}\,\frac{15}{32\,\pi^4}\,\frac{1}{m_aT^6}
    \int_{m_a}^\infty d\omega\, \left(\omega\,
    \frac{\omega^2-m_a^2}{\sinh \frac{\omega}{2T}}\right)^2\bigg/
    \log\frac{\sinh\frac{\omega+\sqrt{\omega^2-m_a^2}}{4T}}{\sinh\frac{\omega-\sqrt{\omega^2-m_a^2}}{4T}}.
\end{equation}
We show this result as a function of $T/m_a$ in Fig.~\ref{fig:Lambda}. For $T\ll m_a$ the effective MFP is exponentially suppressed. The interpretation is that we have defined it to describe energy transport relative to a massless boson and for large $m_a$ relative to $T$, the production of thermal bosons is suppressed.

To estimate the scale for the MFP required to have a significant impact on SN physics, we consider the temperature profile of our numerical SN model shown in Fig.~\ref{fig:SN-Model}. Around a radius of 10\,km the temperature is around 30\,MeV and the temperature gradient 4\,MeV/km, then Eq.~\eqref{fig:Diffuse-Flux} implies a luminosity carried by ALPs of
$L_a\simeq (\lambda_{\rm eff}/{\rm km})\,66\,L_\nu$, where $L_\nu=5.68\times10^{52}\,{\rm erg}/{\rm s}$ is the neutrino luminosity of this model. In other words, unless $\lambda_{\rm eff}\ll 1\,{\rm km}$, ALPs dominate the energy transport within the SN core. On the other hand, for a sufficiently large ALP mass, the effect is much smaller near the PNS surface where temperatures are much smaller.

\begin{figure}[ht]
\centering
\includegraphics[width=0.4\textwidth]{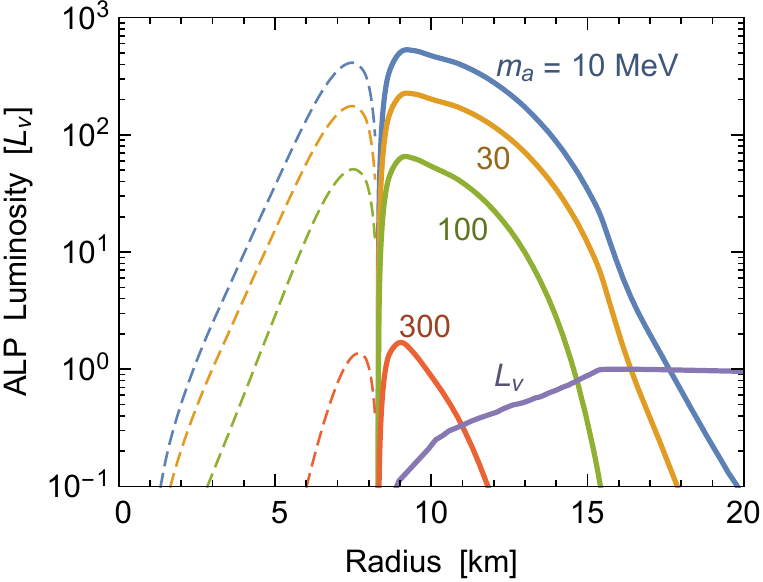}
\caption{Diffusive ALP energy flux carried within our numerical SN model shown in Fig.~\ref{fig:SN-Model}, based on $\lambda_{\rm eff}$ given in Eq.~\eqref{eq:lambda-eff} with $\Gamma_a=1\,{\rm km}^{-1}$ and for the masses $m_a=10$, 30, 100, and 300\,MeV (top to bottom). Negative fluxes (inward bound) shown as dashed lines. We also show the neutrino luminosity $L_\nu(r)$ that reaches its final value between 15 and 16~km. Notice that the luminosities are in local variables, not for a distant observer, and they are in units of $5.68\times10^{52}\,{\rm erg/s}$, the local $L_\nu$ near the decoupling radius.}\label{fig:DiffusiveLa}
\end{figure}

We illustrate this point in Fig.~\ref{fig:DiffusiveLa}, where we show the diffusive ALP flux for $\Gamma_a^{-1}=1\,{\rm km}$ for the indicated range of masses. For small radii, where the $T$ gradient is inward, the negative fluxes are shown as dashed lines. Taking the neutrino decoupling region to be around 17\,km, we see that for $m_a\gtrsim30\,{\rm MeV}$, the ALP flux near the surface is smaller than $L_\nu$, whereas inside it is much larger. To avoid ALPs to dominate energy transfer within the entire SN core, and taking $m_a=100\,{\rm MeV}$, would require $\Gamma_a^{-1}\lesssim 0.01\,{\rm km}$ and thus $G_{a\gamma\gamma}\gtrsim 2\times10^{-6}\,{\rm GeV}^{-1}$.

In Ref.~\cite{Lucente:2020whw} an ALP exclusion plot is shown in the plane of $m_a$ and $G_{a\gamma\gamma}$, where our region of parameters is allowed. Therefore, there is a range of nominally allowed ALP parameters where they would contribute dominantly to energy transfer within the SN core, but not to energy loss. If this effect would actually make an observational difference is another question, but probably it would modify the appearance of convection in the PNS as well as the duration of the neutrino burst. 
In any event, we here have an explicit example of a particle that is too heavy and too short-lived to provide a SN energy-loss channel, yet has a significant effect for the energy transfer within the SN core.

\section{Conclusions}

Motivated by several recent studies about the role of feebly-interacting bosons in stars, notably in supernova cores, we have derived the equations for radiative transfer from first principles for such particles. The main simplification compared with photons is motivated by the feebleness of the interaction and leads us to neglect scattering. So we only consider boson emission and absorption by the background medium. We include systematically the effect of the boson mass that may be comparable to the local temperature or even much larger. After solving the Boltzmann collision equation for a single ray of the boson radiation field, solutions for plane-parallel and spherical geometry follow essentially from phase-space integrations, although these are not entirely trivial.

For the case of spherical geometry, the monochromatic boson luminosity at a radius $r$ from the center of the star is expressed in the form
\begin{equation}\label{eq:finaleq}
    L_\omega(r)=\int_0^\infty dr'\,4\pi r^{\prime\,2} Q_\omega(r')\,{\sf E}_\omega(r,r'),
\end{equation}
where $Q_\omega(r)$ is the monochromatic energy-loss rate for the medium conditions at radius $r$ and ${\sf E}_\omega(r,r')$ is an integral kernel that depends on the reduced boson absorption rate as a function of $r$ or equivalently, the corresponding MFP $\lambda_\omega(r)$. One of our main technical results is to provide the integral kernel explicitly.

The luminosity at a given radius depends on $Q_\omega$ a few MFPs upstream and downstream. If this distance is short compared with the radius itself, the energy flux can be understood in the plane-parallel approximation. In this case the integral kernel simplifies considerably and corresponds to standard results in the literature. Moreover, when the MFP is small compared with the scale height of the temperature variation, one obtains the usual diffusion-limit result, where the energy flux is proportional to the MFP and the temperature gradient. 

Our discussion applies to a stationary situation, when dynamical time scales are long compared with the time it takes for the FIB flux to relax to a stationary solution. In other words, we assumed the FIB flux could be calculated on the basis of a prescribed stellar model without feedback effects. Calculating the FIB flux is then a matter of integrating Eq.~\eqref{eq:finaleq} over the stellar model for every energy $\omega$ and then computing the overall luminosity as an energy integral if the total luminosity is the desired quantity.

In the trapping limit, the contributing region (for every $\omega$) is from a few optical depths to the surface and in this sense a volume integral of significant geometrical extent, not a thin shell near some hypothetical decoupling sphere. We have explicitly studied this question for the case of a ``grey atmosphere,'' where the reduced absorption rate does not depend on energy. On the other hand, the emerging flux is surprisingly well accounted for by assuming it is emitted by a surface at optical depth $\tau=2/3$ with a flux given by the Stefan-Boltzmann law for a blackbody surface corresponding to the radius at $\tau=2/3$ with the local temperature of the background medium. The agreement is best if the temperature varies with a power law $\tau^{1/4}$ as a function of optical depth. The Stefan-Boltzmann recipe has been often used and is surprisingly accurate.

When the reduced absorption rate depends on energy, possibly involving strong variations due to resonance effects, one could apply this approximation separately for every energy $\omega$, but we have not studied how well it approximates the full integration. Of course, the Stefan-Boltzmann approximation is mostly useful as a quick estimate to avoid multi-dimensional numerical integrations that can become cumbersome. However, for the correct result one should simply perform the full volume integration based on the integral kernels that we have provided.

\exclude{In the trapping limit, the boson luminosity can be seen as emerging from a quasi-thermal emission surface at an optical depth $\tau\simeq2/3$. On the other hand, the bosons still emerge from a shell, not a surface, and thus from a volume of considerable radial extent. We clarify the relation between the two perspectives and also find that the picture of quasi-thermal emission from a surface provides an excellent approximation in practice.}

While our derivations and discussions are based entirely on standard radiative transfer theory, not all of our results can be found explicitly in the literature. In this sense we hope that our systematic exposition is useful to the astroparticle community and clarifies some issues that have emerged in the recent literature on FIB emission from stellar bodies.

\section*{Acknowledgements}

We thank Hans-Thomas Janka for helpful discussions on different aspects of this work. AC is supported by the Foreign Postdoctoral Fellowship Program of the Israel Academy of Sciences and Humanities and also acknowledges support from the Israel Science Foundation (Grant 1302/19), the US-Israeli BSF (Grant 2018236), the German-Israeli GIF (Grant I-2524-303.7) and the European Research Council (ERC) under the EU Horizon 2020 Programme (ERC-CoG-2015-Proposal n. 682676 LDMThExp). GR acknowledges support by the German Research Foundation (DFG) through the Collaborative Research Centre “Neutrinos and Dark Matter in Astro and Particle Physics (NDM),” Grant SFB-1258, and under Germany’s Excellence Strategy through the Cluster of Excellence ORIGINS EXC-2094-390783311. EV thanks the Niels Bohr Institute for hospitality, and acknowledges support by the US Department of Energy (DOE) Grant DE-SC0009937, the Rosenfeld Foundation, and the Carlsberg Foundation (CF18-0183).

\bibliographystyle{bibi}
\bibliography{biblio}

\providecommand{\href}[2]{#2}\begingroup\raggedright\begin{thebibliography}{10}

\bibitem{Mikaelian:1978jg}
K.~O. Mikaelian, \emph{{Astrophysical implications of new light Higgs bosons}},
  \href{https://doi.org/10.1103/PhysRevD.18.3605}{\emph{Phys. Rev. D}
  {\bfseries 18} (1978) 3605}.

\bibitem{Dicus:1978fp}
D.~A. Dicus, E.~W. Kolb, V.~L. Teplitz and R.~V. Wagoner, \emph{{Astrophysical
  bounds on the masses of axions and Higgs Particles}},
  \href{https://doi.org/10.1103/PhysRevD.18.1829}{\emph{Phys. Rev. D}
  {\bfseries 18} (1978) 1829}.

\bibitem{Vysotsky:1978dc}
M.~I. Vysotsky, {\relax Ya}.~B. Zel'dovich, M.~{\relax Yu}. Khlopov and V.~M.
  Chechetkin, \emph{{Some astrophysical limitations on the axion mass}},
  {\emph{Pis'ma Zh. Eksp. Teor. Fiz.} {\bfseries 27} (1978) 533}. English
  translation \href{http://jetpletters.ru/ps/1552/article_23764.pdf}{{\em JETP
  Lett.} {\bf 27} (1978) 502}.

\bibitem{Sato:1978vy}
K.~Sato, \emph{{Astrophysical constraints on the axion mass and the number of
  quark flavors}}, \href{https://doi.org/10.1143/PTP.60.1942}{\emph{Prog.
  Theor. Phys.} {\bfseries 60} (1978) 1942}.

\bibitem{Raffelt:1996wa}
G.~G. Raffelt, \emph{{Stars as Laboratories for Fundamental Physics}}.
  University of Chicago Press, 1996.

\bibitem{Chang:2016ntp}
J.~H. Chang, R.~Essig and S.~D. McDermott, \emph{{Revisiting Supernova 1987A
  Constraints on Dark Photons}},
  \href{https://doi.org/10.1007/JHEP01(2017)107}{\emph{JHEP} {\bfseries 01}
  (2017) 107} [\href{https://arxiv.org/abs/1611.03864}{{\ttfamily
  1611.03864}}].

\bibitem{Chang:2018rso}
J.~H. Chang, R.~Essig and S.~D. McDermott, \emph{{Supernova 1987A Constraints
  on Sub-GeV Dark Sectors, Millicharged Particles, the QCD Axion, and an
  Axion-like Particle}},
  \href{https://doi.org/10.1007/JHEP09(2018)051}{\emph{JHEP} {\bfseries 09}
  (2018) 051} [\href{https://arxiv.org/abs/1803.00993}{{\ttfamily
  1803.00993}}].

\bibitem{Lucente:2020whw}
G.~Lucente, P.~Carenza, T.~Fischer, M.~Giannotti and A.~Mirizzi, \emph{{Heavy
  axion-like particles and core-collapse supernovae: constraints and impact on
  the explosion mechanism}},
  \href{https://doi.org/10.1088/1475-7516/2020/12/008}{\emph{JCAP} {\bfseries
  12} (2020) 008} [\href{https://arxiv.org/abs/2008.04918}{{\ttfamily
  2008.04918}}].

\bibitem{Bollig:2020xdr}
R.~Bollig, W.~DeRocco, P.~W. Graham and H.-T. Janka, \emph{{Muons in
  supernovae: implications for the axion-muon coupling}},
  \href{https://doi.org/10.1103/PhysRevLett.125.051104}{\emph{Phys. Rev. Lett.}
  {\bfseries 125} (2020) 051104}
  [\href{https://arxiv.org/abs/2005.07141}{{\ttfamily 2005.07141}}].

\bibitem{Caputo:2021rux}
A.~Caputo, G.~Raffelt and E.~Vitagliano, \emph{{Muonic boson limits: Supernova
  redux}}, \href{https://doi.org/10.1103/PhysRevD.105.035022}{\emph{Phys. Rev.
  D} {\bfseries 105} (2022) 035022}
  [\href{https://arxiv.org/abs/2109.03244}{{\ttfamily 2109.03244}}].

\bibitem{Caputo:2022mah}
A.~Caputo, H.-T. Janka, G.~Raffelt and E.~Vitagliano, \emph{{Low-Energy
  Supernovae Severely Constrain Radiative Particle Decays}},
  \href{https://arxiv.org/abs/2201.09890}{{\ttfamily 2201.09890}}.

\bibitem{Croon:2020lrf}
D.~Croon, G.~Elor, R.~K. Leane and S.~D. McDermott, \emph{{Supernova muons: New
  constraints on $Z'$ bosons, axions and ALPs}},
  \href{https://doi.org/10.1007/JHEP01(2021)107}{\emph{JHEP} {\bfseries 01}
  (2021) 107} [\href{https://arxiv.org/abs/2006.13942}{{\ttfamily
  2006.13942}}].

\bibitem{Burrows:1990pk}
A.~Burrows, M.~T. Ressell and M.~S. Turner, \emph{{Axions and SN1987A: Axion
  trapping}}, \href{https://doi.org/10.1103/PhysRevD.42.3297}{\emph{Phys. Rev.
  D} {\bfseries 42} (1990) 3297}.

\bibitem{Raffelt:1988rx}
G.~G. Raffelt and G.~D. Starkman, \emph{{Stellar energy transfer by keV-mass
  scalars}}, \href{https://doi.org/10.1103/PhysRevD.40.942}{\emph{Phys. Rev. D}
  {\bfseries 40} (1989) 942}.

\bibitem{Rutten:2003}
R.~J. {Rutten}, \emph{{Radiative Transfer in Stellar Atmospheres}},  2003.
\newblock Utrecht University Lecture Notes,
  \url{https://robrutten.nl/rrweb/rjr-pubs/2003rtsa.book.....R.pdf}.

\bibitem{Mihalas:1978}
D.~{Mihalas}, \emph{{Stellar atmospheres (2nd ed.)}}. Freeman, 1978.

\bibitem{Shapiro:1983du}
S.~Shapiro and S.~Teukolsky, \emph{{Black holes, white dwarfs, and neutron
  stars: The physics of compact objects}}. John Wiley \& Sons, 1983.

\bibitem{ThorneBook}
K.~S. {Thorne} and R.~D. {Blandford}, \emph{{Modern Classical Physics: Optics,
  Fluids, Plasmas, Elasticity, Relativity, and Statistical Physics}}. Princeton
  University Press, 2017.

\bibitem{ThorneWeb}
\emph{Applications of classical physics},
  \url{http://www.pmaweb.caltech.edu/Courses/ph136/yr2012/}.

\bibitem{Lucente:2022wai}
G.~Lucente, O.~Straniero, P.~Carenza, M.~Giannotti and A.~Mirizzi,
  \emph{{Constraining heavy axion-like particles by energy deposition in
  Globular Cluster stars}},  \href{https://arxiv.org/abs/2203.01336}{{\ttfamily
  2203.01336}}.

\bibitem{Weldon:1983jn}
H.~A. Weldon, \emph{{Simple rules for discontinuities in finite temperature
  field theory}}, \href{https://doi.org/10.1103/PhysRevD.28.2007}{\emph{Phys.
  Rev. D} {\bfseries 28} (1983) 2007}.

\bibitem{CCSNarchive}
\emph{Garching core-collapse supernova research archive},
  \url{https://wwwmpa.mpa-garching.mpg.de/ccsnarchive/}.

\bibitem{Kourganoff:1952}
V.~{Kourganoff}, \emph{{Basic methods in transfer problems; radiative
  equilibrium and neutron diffusion}}. Clarendon Press, Oxford, 1952.

\bibitem{Raffelt:1985nk}
G.~G. Raffelt, \emph{{Astrophysical axion bounds diminished by screening
  effects}}, \href{https://doi.org/10.1103/PhysRevD.33.897}{\emph{Phys. Rev. D}
  {\bfseries 33} (1986) 897}.

\end{thebibliography}\endgroup

\end{document}